\documentclass[fleqn,usenatbib]{mnras}

\usepackage{newtxtext,newtxmath}

\usepackage[T1]{fontenc}

\DeclareRobustCommand{\VAN}[3]{#2}
\let\VANthebibliography\thebibliography
\def\thebibliography{\DeclareRobustCommand{\VAN}[3]{##3}\VANthebibliography}

\usepackage{graphicx}	
\usepackage{amsmath}	
\usepackage{subcaption}
\usepackage{caption}
\usepackage{longtable}
\usepackage{multirow,tabularx}
\usepackage{color}
\UseRawInputEncoding

\title[Coronal lines in qiz]{Delayed Appearance and Evolution of Coronal Lines in the TDE AT2019qiz}

\author[P. Short et al.]{
P.~Short$^{1}$\thanks{E-mail: pshort@roe.ac.uk},
A.~Lawrence$^{1}$,
M.~Nicholl$^{2}$,
M.~Ward$^{3}$,
T.~M.~Reynolds$^{4}$,
S.~Mattila,$^{5,20}$
C.~Yin$^{1}$,
I.~Arcavi$^{6,7}$,
\newauthor 
A.~Carnall$^{1}$,
P.~Charalampopoulos$^{8}$,
M.~Gromadzki$^{9}$,
P.~G.~Jonker$^{10,11}$,
S.~Kim$^{12,13}$,
G.~Leloudas$^{8}$,
\newauthor 
I.~Mandel$^{14,15}$,
F.~Onori$^{16}$,
M.~Pursiainen$^{8}$,
S.~Schulze$^{17}$,
C.~Villforth$^{18}$,
T.~Wevers$^{19}$
\\
$^{1}$Institute for Astronomy, University of Edinburgh, Royal Observatory, Blackford Hill, Edinburgh EH9 3HJ, UK\\
$^{2}$Birmingham Institute for Gravitational Wave Astronomy and School of Physics and Astronomy, University of Birmingham, Birmingham, B15 2TT, UK\\
$^{3}$Centre for Extragalactic Astronomy, Department of Physics, University of Durham, South Road, Durham, DH1 3LE, UK\\
$^{4}$Cosmic Dawn Center (DAWN), Niels Bohr Institute, University of Copenhagen, Denmark\\
$^{5}$Tuorla Observatory, Department of Physics and Astronomy, University of Turku, FI-20014 Turku, Finland\\
$^{6}$The School of Physics and Astronomy, Tel Aviv University, Tel Aviv 69978, Israel\\
$^{7}$CIFAR Azrieli Global Scholars program, CIFAR, Toronto, Canada\\
$^{8}$DTU Space, National Space Institute, Technical University of Denmark, Elektrovej 327, 2800, Kgs. Lyngby, Denmark\\
$^{9}$Astronomical Observatory, University of Warsaw, Al. Ujazdowskie 4, 00-478 Warszawa, Poland\\
$^{10}$SRON, Netherlands Institute for Space Research, Niels Bohrweg 4, 2333 CA, Leiden, The Netherlands\\
$^{11}$Department of Astrophysics/IMAPP, Radboud University Nijmegen, P.O. Box 9010, 6500 GL, Nijmegen, The Netherlands\\
$^{12}$Astro-engineering center (AIUC), Instituto de Astrofísica, Pontificia Universidad Católica de Chile, Santiago, Chile\\
$^{13}$Max-Planck-Institut für Astronomie, Königstuhl 17, 69117 Heidelberg, Germany\\
$^{14}$Monash Centre for Astrophysics, School of Physics and Astronomy, Monash University, Clayton, Victoria 3800, Australia\\
$^{15}$ARC Center of Excellence for Gravitational Wave Discovery — OzGrav, Australia\\
$^{16}$INAF-Osservatorio Astronomico d'Abruzzo, via M. Maggini snc, I-64100 Teramo, Italy\\
$^{17}$The Oskar Klein Centre, Department of Physics, Stockholm University, AlbaNova, SE-106 91 Stockholm, Sweden\\
$^{18}$University of Bath, Department of Physics, Claverton Down, Bath, BA2 7AY, UK\\
$^{19}$European Southern Observatory, Alonso de Córdova 3107, Casilla 19, Santiago, Chile\\
$^{20}$School of Sciences, European University Cyprus, Diogenes street, Engomi, 1516 Nicosia, Cyprus
}

\begin{document}
\label{firstpage}
\pagerange{\pageref{firstpage}--\pageref{lastpage}}
\maketitle

\begin{abstract}

Tidal disruption events (TDEs) occur when a star gets torn apart by a supermassive black hole as it crosses its tidal radius. We present late-time optical and X-ray observations of the nuclear transient AT2019qiz,  which showed the typical signs of an optical-UV transient class commonly believed to be TDEs. Optical spectra were obtained 428, 481 and 828 rest-frame days after optical lightcurve peak, and a UV/X-ray observation coincided with the later spectrum. The optical spectra show strong coronal emission lines, including [Fe\,VII], [Fe\,X], [Fe\,XI] and [Fe\,XIV]. The Fe lines rise and then fall, except [Fe\,XIV] which appears late and rises. We observe increasing flux of narrow H$\alpha$ and H$\beta$ and a decrease in broad H$\alpha$ flux. The coronal lines have FWHMs ranging from $\sim150-300$km\,s$^{-1}$, suggesting they originate from a region between the broad and narrow line emitting gas. Between the optical flare and late-time observation, the X-ray spectrum softens dramatically. The 0.3-1\,keV X-ray flux increases by a factor of $\sim50$ while the hard X-ray flux decreases by a factor of $\sim6$. WISE fluxes also rose over the same period, indicating the presence of an infrared echo. With AT2017gge, AT2019qiz is one of two examples of a spectroscopically-confirmed optical-UV TDE showing delayed coronal line emission, supporting speculations that Extreme Coronal Line Emitters in quiescent galaxies can be echos of unobserved past TDEs. We argue that the coronal lines, narrow lines, and infrared emission arise from the illumination of pre-existing material likely related to either a previous TDE or AGN activity.

\end{abstract}




\section{Introduction} \label{intro}

\begin{figure*}
\centering
\includegraphics[width=1\linewidth]{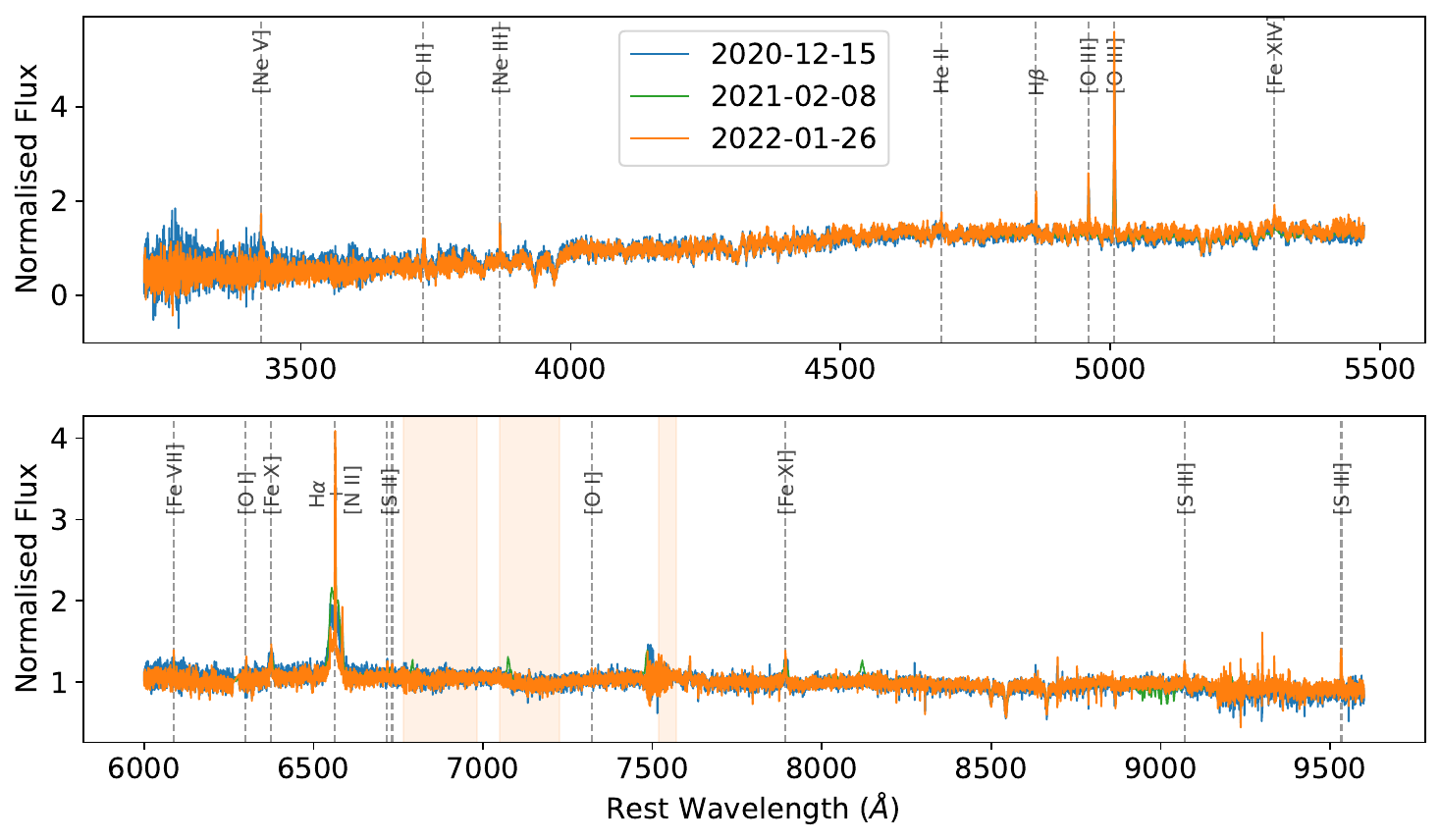}
\caption{The reduced X-Shooter and MUSE spectra. The X-Shooter spectra are binned for ease of viewing. Emission line identifications are marked with the grey dashed lines. The upper plot covers the X-Shooter UVB arm wavelength range while the lower plot covers the VIS arm. The `emission' features redward of $\sim$6800\AA{} in the MUSE (green) spectrum are residuals from telluric corrections. The fluxes are scaled to approximately match the stellar continuum in each case.}
\label{fig:specs}
\end{figure*}

A tidal disruption event (TDE) occurs when a star's orbit brings it within the tidal radius of its host galaxy's central supermassive black hole (SMBH). The star is subsequently ripped apart, releasing a luminous flare as the stellar material falls back onto the SMBH \citep{hills75, rees88}. Early TDE candidates were discovered via soft X-ray flares (e.g. \citealt{kom99, komgren}), with luminosities of $\sim10^{44}$\,erg\,s$^{-1}$ and lightcurves matching the $t^{-5/3}$ decay rate expected from theory \citep{hills75, rees88}. Since then, a variety of nuclear transients and extreme variables has been observed, but a consistent class of optical-UV transients has emerged which are widely accepted as being TDEs \citep{vanVelzen20, gezari21}. The key features which define this class include (i) a large amplitude nuclear optical flare; (ii) a rise time of $\sim$days-weeks and a decay time of $\sim$weeks-months; (iii) extremely broad and weak emission lines from Hydrogen and Helium; (iv) a blue continuum with effective temperature around 10-30,000K; and (v) roughly constant optical colours during the decay. We refer to members of this class as ``spectroscopically confirmed optical-UV TDEs''. Strictly speaking it is not rigorously proved that these events result from the disruption of a star, but most workers accept that a TDE is the most likely explanation, and so refer to these objects simply as ``TDEs'' rather than for example ``candidate TDEs''. Empirically they form a well defined class which is distinct from either known types of supernovae (SNe), or extreme active galactic nuclei (AGN) variables such as Changing Look AGN. 

Even within this consistent class, there is a detailed diversity. The appearance or lack of Hydrogen, Helium and Bowen fluorescence emission lines divide these optically-selected TDEs into three classes; TDE-H, TDE-He and TDE-H+He \citep{arcavi14, lelodas19, vv21}. It has been shown that the strengths of these lines can vary as the TDE evolves (e.g. \citealt{nicholl17}, \citealt{panos22}, \citealt{onori22}), allowing TDEs to transition between classes. In addition, other features have been observed such as Fe\,II lines \citep{wevers192,cann20} and double-peaked emission lines \citep{short20,hung20,wevers22}. TDEs are also extremely diverse in their X-ray to optical ratios, ranging from $\sim10^{-4}$ to 1 \citep{auch17}. This could indicate that some TDEs form accretion disks quickly while others do not, or that X-rays are heavily absorbed/re-processed by an atmosphere or outflow along the observer's line of sight (e.g. \citealt{loeb97}, \citealt{guill14}, \citealt{roth16}, \citealt{roth18}, \citealt{dai18}).\\

Given that optical-UV TDEs are seen to occur in systems with black hole masses $10^{6-7} M\odot$, \citep{wevers17,mosfit22} one would expect a compact accretion disc to have a temperature $T\sim 10^6$~K, at odds with the observed temperatures of $T\sim 2-4 \times 10^4$~K. This may be explained by reprocessing (see paragraph above), but is the expected EUV continuum actually there? One test is to look for emission from highly ionised species.\\

Prior to the current era of optical sky surveys, \cite{kom08} discovered fading high-ionisation coronal Iron emission lines, He\,II and double-peaked Balmer lines in the nearby galaxy SDSS J095209.56+214313.3 (J0952+2143) during a search for emission line galaxies in SDSS DR6. Optical photometry showed a peak in optical brightness in 2004, with an increase towards the NIR which was inconsistent with previous 2MASS observations. GALEX data also showed a rise in UV flux. \cite{wang11} found a similar example of coronal line emission in the galaxy SDSS J074820.67+471214.3 which also had an optical flair, and went on to find 5 more in a systematic search for such objects \citep{wang12}. These authors proposed that the most likely cause of such coronal emission was an earlier, unobserved, TDE in a gas rich environment. Of the five new events reported by \citealt{wang12}, three had fading iron lines as found by \citep{yang13}, making these and J0952+2143 the strongest candidates for ECLEs which are caused by illumination by an earlier TDE. The case for J0952+2143 was made stronger, when a clear infrared echo was found \cite{kom09}, and when the associated optical flare was covered more densely in LINEAR data \citep{palaversa16} retrospectively.\\

In this scenario the broad low-ionisation emission lines come from re-processing of UV/X-ray flares by outflows directly connected with the TDE in question, which matches well the line profiles of H$\alpha$ and He\,II \citep{roth18}, while the coronal lines originate from interactions with a clumpy interstellar medium or other pre-existing material. If produced by photoionisation, the coronal lines imply an ionising continuum extending into the X-rays, requiring ionising potentials of up to nearly 400eV in the case of [Fe\,XIV]. Coronal lines are not uncommon in AGN \citep{grandi78, penston84, gelbord09}, where we also expect a hard ionising continuum.\\

It is therefore important to test for the presence of highly ionised species in the class of well-confirmed optical-UV TDEs, where we have high quality optical, X-ray and spectroscopic data at the time of the actual flare. Coronal lines have so far only been observed in one such clear optical-UV TDE to date, AT2017gge \citep{onori22}. The lines appeared in this event from $\sim200-1700$ days after the optical lightcurve peak. Another potential candidate for a TDE showing coronal lines is the nuclear transient AT\,2019avd. \cite{avd} observed [Fe\,X] and [Fe\,XI] emission lines accompanied by a rise in X-ray flux, and \cite{Chen2022} argue that this event could be interpreted as a TDE. However observationally it is definitely not a member of our well defined class of optical-UV TDEs - it has a double peaked light curve, and relatively strong and narrow Balmer lines. There was also a tentative detection of coronal lines in the nuclear transient PS16dtm (Petrushevska et al 2023). Like AT\,2019avd, it does not show all the classic signs of the optical-UV TDE class, but is nonetheless argued to be a TDE by \cite{petrushevska23}.\\

As well as exhibiting coronal lines, AT2017gge was also exceptionally luminous in the infrared (IR). This IR emission arises from an IR echo, in which dust in the vicinity of the SMBH absorbs UV/optical emission produced by the TDE, and re-radiates this energy in the IR. IR echoes have previously been observed in approximately half of optically discovered TDEs, but typically the total proportion of luminosity arising from the dust is small, implying that the covering factor is only $\sim$1\% of that observed in the UV/optical \citep{Jiang2021b}. The covering factor derived for the IR echo associated with AT2017gge was $\sim$20\% \citep{Wang2022}, much larger than typical TDEs discovered in optical and occurring in quiescent galaxies, and indicating the presence of a larger quantity of dust. AT2019avd also exhibited an abnormally bright IR echo \citep{avd,Chen2022}, further suggesting a connection between these phenomena.  \\

\begin{figure*}
\centering
\includegraphics[width=1\linewidth]{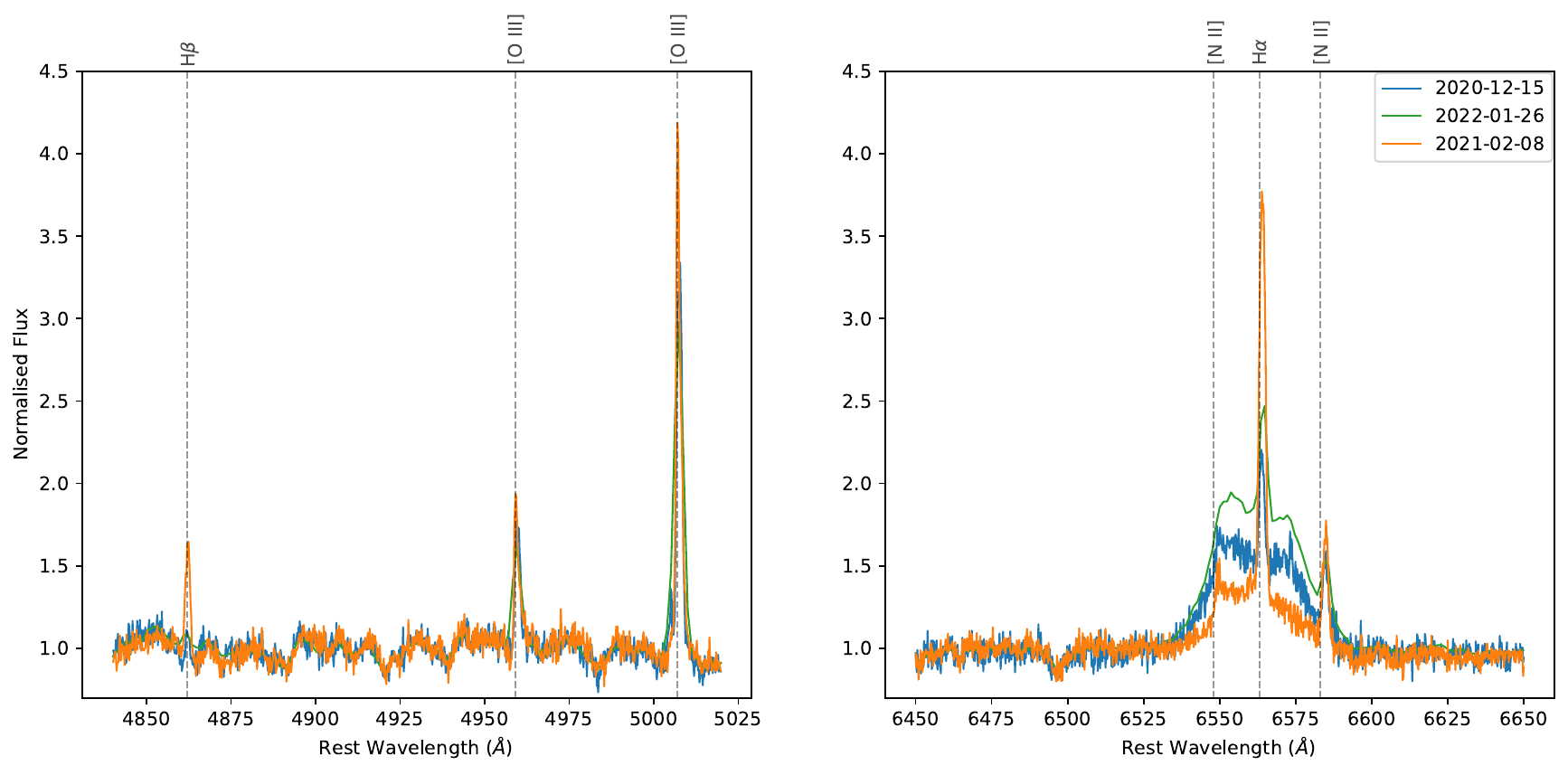}
\caption{The H$\beta$ + [O III] and H$\alpha$ regions for our late-time spectra. The narrow Balmer lines have brightened considerably over time, while the broad H$\alpha$ has risen in the Feb 2021 (green) spectrum before dimming again in the Jan 2022 spectrum (orange).}
\label{fig:balmer}
\end{figure*} 

The subject of this work is AT\,2019qiz, a clear member of the optical-UV TDE class first identified on 2019-09-19 \citep{atel} at a redshift of $z=0.0153$, and reaching a peak at 2019-10-10. \textit{Gaia} data (Gaia Science Alerts \citealt{gsa}) show no further activity in the optical lightcurve since the decline from this peak. The transient was analysed in detail by \cite{qiz20} and \cite{qiz21}. Lightcurve fitting results suggested that the flare was produced by a $\sim$1M$_{\odot}$ star being disrupted by a $\sim$10$^{6}$M$_{\odot}$ black hole. The black hole mass derived from the TDE light curve was consistent with that derived from velocity dispersion measurements and $M-\sigma$ relations. Early optical spectra show broad, asymmetric H and He\,II lines, which \cite{qiz20} argued were driven by an expanding outflow. The lines became more symmetrical as the TDE evolved. He\,II was replaced by N\,III\,$\lambda4641$ via Bowen fluorescence \citep{qiz20, qiz21}, which occurs when recombining He\,II produces an EUV photon which goes on to excite some O\,III and N\,III states. These subsequently recombine producing `Bowen fluorescence' lines. UV spectra show high-ionisation broad absorption lines (HiBALs) and Fe and low-ionisation broad absorption lines (FeLoBALs) \citep{qiz21}. Unfortunately, there are no spectra of the host galaxy (2MASX J04463790-1013349) taken before the flare. Both studies suggest that the host galaxy harbours a weak AGN based on Baldwin-Phillips-Terlevich (BPT) diagrams. Early X-ray detections suggest accretion started promptly in this event \citep{qiz20}, though the high hardness ratio may indicate that the X-ray emission originates from the pre-existing AGN \citep{qiz21}. However, \cite{qiz20} note that the hardness ratio varied during outburst, implying that the TDE did affect the X-ray emission.\\

In this paper we present and analyse two new spectra of AT2019qiz obtained with X-Shooter taken 428 and 828 rest-frame days after the optical lightcurve peak, a MUSE spectrum obtained 481 rest-frame days after optical peak and \textit{Swift} XRT data taken 816 rest-frame days post optical peak, as well as re-examining the spectra taken during the campaign of \cite{qiz20}. In \S\ref{obs} we detail our observations and data reduction process. In \S\ref{spec} we present our spectroscopic analysis, in \S\ref{swift} we present analysis of \textit{Swift} UV and X-ray data, and in \S\ref{sed} we construct an SED of the TDE in outburst. In \S\ref{IR} we present analysis of NEOWISE IR data. In \S\ref{disc} we discuss and interpret our results and analysis and in \S\ref{conc} we provide a summary and conclusion to this work.

\section{Observations and Data Reduction} \label{obs}
\subsection{X-Shooter}
Observations were made on 2020-12-15 and 2022-01-26 with the X-Shooter instrument \citep{xsho} mounted on UT3 at the Very Large Telescope (VLT) in ESO's Paranal observatory, Chile. X-Shooter is an intermediate resolving power spectrograph with a wavelength range spanning 3000 - 25,000\AA{}. The spectrograph consists of three arms; the UVB arm which covers 3000-5595\AA{}, the VIS arm which spans 5595-10,240\AA{} and the NIR arm which ranges from 10,240-24,800\AA{}. Slit widths of 1.0", 0.9" and 0.9" were used for the UVB, VIS and NIR arms respectively, giving respective resolving powers of R=5400, 8900 and 5600. Data were obtained under programs 106.21SS.001 (PI Short) and 108.22J7.001 (PI Nicholl). We refer to these spectra throughout this paper as the Dec 2020 and Jan 2022 spectra, respectively.\\

Data reduction was performed using the X-Shooter pipeline recipes and the EsoReflex GUI environment \citep{esor}. The NIR arm was reduced in NOD mode while the UVB and VIS arms were reduced in STARE mode which was found to improve the signal-to-noise. Telluric corrections were performed in the VIS arm using telluric standards observed before and after the observations at a similar position to the target. Telluric features in the NIR arm proved difficult to correct and, as this wavelength region contained no features of interest, this arm was not used in the rest of our analysis. Extinction correction was performed using the \cite{F99} model in the \textsc{Extinction}\footnote{\href{https://extinction.readthedocs.io/en/latest/}{https://extinction.readthedocs.io/en/latest/}} package in Python, with reddening values from the NASA/IPAC Infrared Science Archive \citep{ebv}. As the optical continuum luminosity at this phase is dominated by the host galaxy, the spectra were flux corrected using the \textsc{PySynphot} Python package \citep{synp} to an aperture-matched host \textit{r}-band mag of 16.47 as determined in \cite{qiz20}. The reduced spectra are displayed in Figure \ref{fig:specs}.\\

\subsection{MUSE}
We observed the field on 2021-02-08 with the panoramic integral-field spectrograph MUSE \citep{bacon2010}, mounted at UT4 at ESO's Very Large Telescope, in the seeing-enhanced adaptive optics wide-field mode (WFM-AO) under clear condition and DIMM seeing $<$ 0.6$^{\prime\prime}$. MUSE WFM-AO has a large field of view covering $1'\times1'$ and a high spatial sampling of $0.2''\times0.2''$. It covers the wavelength range from 4650 to 9300~\AA{} with a spectral resolving power of 2000--4000. The observation consists of three 1100-s exposures that were dithered by 1--$2''$ and rotated by $90^\circ$ with respect to each other. We retrieved the reduced science-ready datacube from the ESO archive, which was reduced with the ESO MUSE pipeline version 2.8.4 \citep{weilbacher2020}. We extracted the spectrum of the host galaxy nucleus using a circular aperture with a diameter of $1\times$FWHM(stellar PSF). The integral field of MUSE also allows us to check for spatial extent. The host starlight is of course very extended, but the emission line source is compact. Comparing extractions with 1 arcsec and 2 arcsec apertures changed the [OIII] flux by $\sim$15\%. Given the other uncertainties, this gives us reasonable confidence in both the Xshooter and MUSE line flux values.

\begin{figure*}
\centering
\includegraphics[width=1\linewidth]{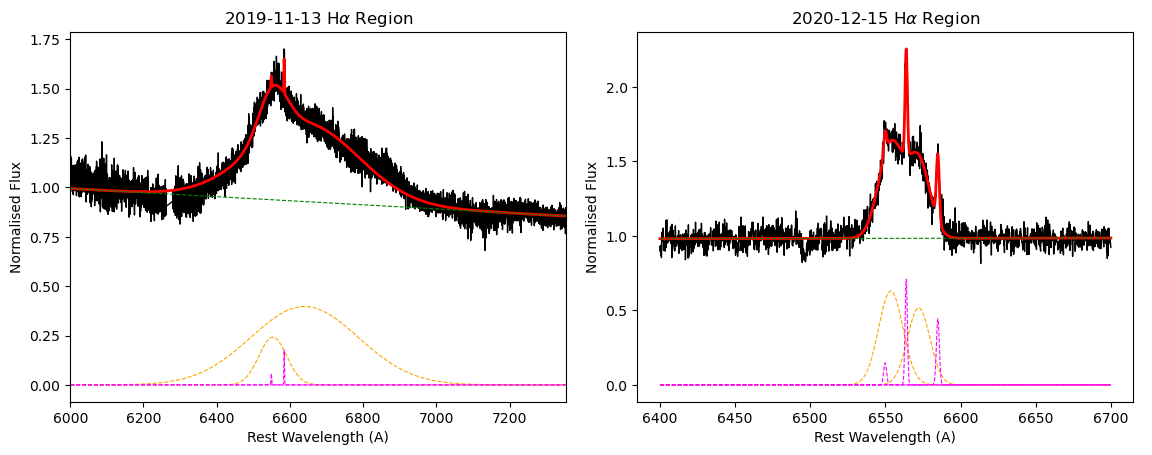}
\caption{Fits to H$\alpha$ regions in (left) an early X-Shooter spectra obtained 20191113 and (right) our Dec 2020 spectrum. In both cases the broad H$\alpha$ feature is well fit by two broad Gaussian components, though in the earlier spectra the second component is much broader and offset.}
\label{fig:ha_fits}
\end{figure*}

\begin{figure}
\centering
\includegraphics[width=1\linewidth]{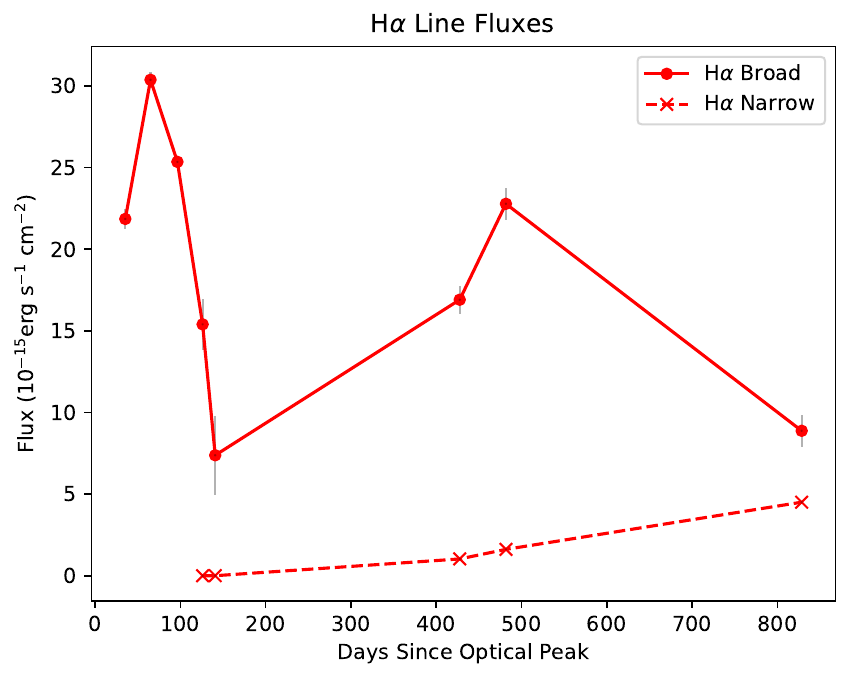}
\caption{The evolution of both the broad and narrow H$\alpha$ emission line components using spectra from \citealt{qiz20} and the spectra presented in this paper. In the first three spectra there is a second broad outflow component which we omit from our measurements. Errorbars are included but in some cases too small to be visible.}
\label{fig:ha_fluxes}
\end{figure}

\subsection{Neil Gehrels Swift Observatory}
A \textit{Swift} \citep{gehrels04} observation was made on 2022-01-13 using the XRT instrument in photon-counting mode and the UVOT instrument with the UVW2 filter (Obs ID: 00012012043, PI Short). The XRT data was reduced and a spectrum was generated using the online \textit{Swift} XRT product builder \citep{swift}. The UVOT UVW2 flux was measured using a 5'' aperture, approximately twice the UVOT point-spread function. This matches the aperture used to extract earlier UVOT photometry by \cite{qiz20}. The count rates were obtained using the Swift \textsc{uvotsource} tools and converted to magnitudes using the UVOT photometric zero points \citep{breev11}. The analysis pipeline used software HEADAS 6.24 and UVOT calibration 20170922. We note that the calibration files have since been updated, but we use the older calibration for consistency with \cite{qiz20}. We measured UVW2 AB magnitude of $20.58\pm0.04$ which is consistent with the host magnitude measured in \citep{qiz20} of $20.51\pm0.2$, suggesting there is no detectable contribution from the TDE.

\subsection{NEOWISE survey}

We obtained MIR data of AT 2019qiz taken as part of the NEOWISE survey \citep{Mainzer2011,Mainzer2014} from the public NEOWISE-R Single Exposure (L1b) Source Table\footnote{\url{https://irsa.ipac.caltech.edu/cgi-bin/Gator/nph-scan?mission=irsa&submit=Select&projshort=WISE}}. NEOWISE observes the entire sky at 6 month intervals, with multiple observations at each "visit". We adopted the median value of the individual measurements at each visit as the magnitude, removing measurements flagged as poor quality or separated from the coordinates of AT 2019qiz by more than 2\arcsec. For the uncertainty, we adopted the standard error of mean of the individual measurements taken at each bi-yearly epoch of WISE observation after 3 sigma-clipping outliers. We added 0.0026 mag and 0.0061 mag uncertainties in quadrature to the W1 and W2 measurements, respectively, which are the RMS residuals found in the photometric calibration during the survey period. No extinction correction was made to the WISE magnitudes.

\section{Spectroscopic Analysis} \label{spec}

\subsection{Host Black Hole Mass}
\cite{qiz20} follow the method of \cite{wevers17} and \cite{wevers191} to fit the velocity dispersion of stellar absorption lines using \textsc{PPXF} \citep{ppxf}. They used an X-Shooter spectrum obtained 140 days after optical lightcurve peak, and measured a velocity dispersion of $70\pm2$km\,s$^{-1}$. We use the same method on our X-Shooter spectra and measure a value of $72\pm1$km\,s$^{-1}$, consistent with that measured in \cite{qiz20}. Using the relations of \cite{mcma13}, \cite{gult09} and \cite{koho13} our velocity dispersion measurement yields black hole masses of log$_{10}(M_{BH}/M_{\odot})$ = $5.82\pm0.41$, $6.24\pm0.48$ and $6.54\pm0.32$, respectively. Both \cite{qiz20} and \cite{qiz21} obtain additional black hole mass estimates from UV and optical lightcurve model fits, deriving values of log$_{10}(M_{BH}/M_{\odot})$ = $5.89^{+0.05}_{-0.06}$ and $6.14\pm0.10$, respectively. These are both consistent with the estimate from the velocity dispersion. Additionally, \cite{mosfit22} re-fit the lightcurve after correcting for a calibration error affecting UVOT photometry at the time of the AT2019qiz outburst. The slightly increased value of log$_{10}(M_{BH}/M_{\odot})$ = 6.22 is still consistent with the velocity dispersion results. Given the range of estimated values, for simplicity we assume an intermediate SMBH mass of $M_{BH}=10^6M_{\odot}$ for the remainder of this analysis, but note that this has an uncertainty of at least a factor of two.

\subsection{Low Ionisation Lines} \label{low-ion}

\begin{figure*}
\centering
\includegraphics[width=1\linewidth]{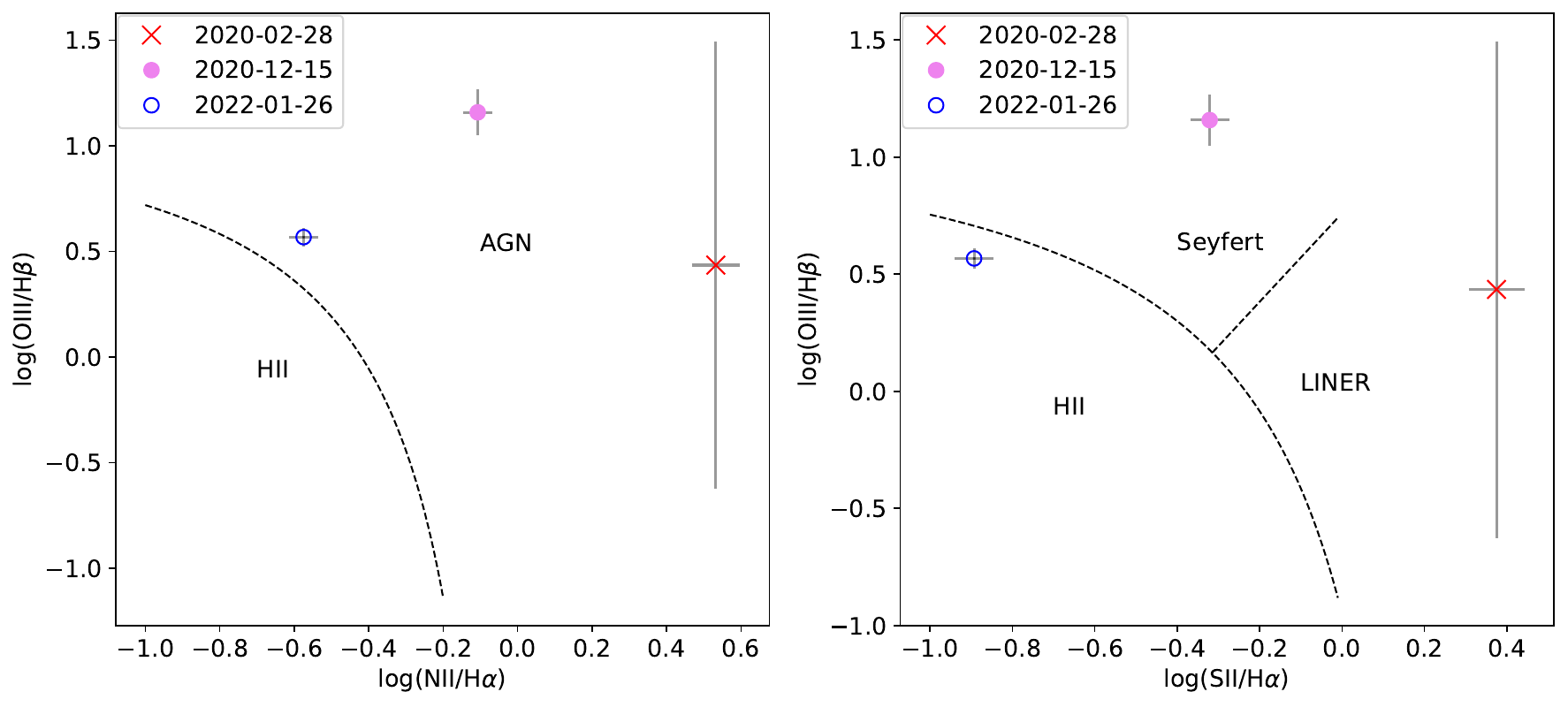}
\caption{BPT diagrams showing the location of the nuclear emission in both our spectra. Measurements from the Dec 2020 spectrum clearly suggest the emission is due to an AGN, but the Jan 2022 emission lines more closely resemble a star forming region. However, H$\beta$ could be overestimated if there is an unresolved component from the TDE.}
\label{fig:bpt}
\end{figure*}

The spectra feature narrow emission lines, including [O\,II], [O\,III], [N\,II], [S\,II], H$\alpha$ and H$\beta$, as well as broad H$\alpha$. The H$\beta$ and H$\alpha$ regions are displayed in Figure \ref{fig:balmer}. To measure the flux and widths of each line we use the Python package \textsc{lmfit}\footnote{\href{https://lmfit.github.io/lmfit-py/}{https://lmfit.github.io/lmfit-py/}} to fit a single Gaussian component to the line and a polynomial component to the local continuum in that line region. In order to separate the narrow H$\alpha$ and [N\,II] emission lines from the broad H$\alpha$ feature, we fit the entire region with 5 Gaussian components simultaneously. The broad feature is not well fit by a single Gaussian so we fit two components to this, while the remaining three components are fit to the narrow H$\alpha$ and [N\,II]\,$\lambda\lambda$\,6549,6584\AA{} lines. The [N\,II]\,$\lambda$6584\AA{}/$\lambda$6549\AA{} ratio is fixed to 3 and the [N\,II]\,$\lambda$6549\AA{} line width is fixed to match that of the 6584\AA{} line. The fits, shown in the right panel of Figure \ref{fig:ha_fits}, were performed on both X-Shooter spectra and the MUSE spectrum and provide a satisfactory fit, hinting that the broad line may be double peaked. When measuring H$\beta$, an absorption component is clearly visible which we fit separately. A stellar absorption feature is likely obscured in the H$\alpha$ blend, meaning that the total H$\alpha$ flux could be underestimated. Flux measurements of all emission lines are shown in tables \ref{tab:fe_flx1}, \ref{tab:fe_flx2} and \ref{tab:fe_flx3}. We note that the third spectrum seems to have an anomalous [OIII] 5007/4959 ratio. However the ratio of peak values is much closer to the expected value of 3. There is an additional systematic error in deriving line fluxes that arises from the uncertain line wings in the presence of both noise and structured stellar continuum. We estimate this systematic error to be $\sim$ 15\% of the line flux.\\

There are a number of changes in the emission lines from the Dec 2020 spectrum to the Jan 2022 spectrum. The broad H$\alpha$ line increases in flux from the Dec 2020 to Feb 2021 spectra but then drops again in the Jan 2022 spectrum. Meanwhile, the narrow H$\alpha$, H$\beta$ and [S\,III] lines increase in strength. The [O\,III] lines do not change significantly. We also observe the appearance of [O\,I]\,$\lambda6300$\AA{} and He\,II\,$\lambda$4686\AA{} in the Jan 2022 spectrum, neither of which were detectable in the previous spectra.\\

In Figure 4 we plot the evolution of both broad and narrow H$\alpha$ emission over time. We use X-Shooter spectra from \cite{qiz20} dating back to $\sim35$ days after optical lightcurve peak to show the complete evolution. In the spectra from \cite{qiz20} there are two broad components but one is consistently broader and redshifted. \cite{qiz20} determine this is possibly due to an outflow so we exclude it from our measurements. An example fit to these earlier spectra is included in Figure 3. The broad line evolution initially follows a downward trend but increases again at some point between $\sim130$ days and 481 days (our Dec 2020 spectrum). It peaks at 481 days (our Feb 2021 spectrum) before declining again. The narrow line emission appears to be increasing. This is presumably due to photons from the initial outburst illuminating material at large radii. The fact that this is taking place gradually over an extended time, rather than a simple spike at a time delayed from the outburst, indicates that the material is spread over a large range of distances and/or angles. We note that an increase in narrow line flux was also observed in the other coronal line emitting TDE (AT2017gge; \citealt{onori22}). In addition,\citet{wang11, wang12}
 see [O III] increase after the coronal lines have faded in the Extreme Coronal Line Emitters which they proposed to be the echoes of TDEs.\\

The fact that the Balmer line increase has not yet come to a clear maximum means that we do not yet have a clear picture of the radial and angular distribution of the related gas. However, we can ask whether what we see so far is consistent with the kinematic information. We have neither a kinematic or spatial model, but a reasonable expectation is that line widths will be of the order of the virial velocity at the corresponding distance, so that we can estimate distance crudely as $r=GM/v_{FWHM}^2$, using a black hole mass of $M=10^6$ M$\odot$. Balmer line widths are $\sim$80km s$^{-1}$ indicating a distance of ~2.2 light years. However at this distance, the gas may be responding to the galaxy potential as much as the black hole, so that this distance should be considered a lower limit. This is consistent with the fact that Balmer line response has not yet peaked; on the other hand there clearly is significant material already responding at delays less than two years, indicating that the illuminated material is distributed over a large range of radii and/or angles.\\

\begin{figure*}
\centering
\includegraphics[width=0.9\linewidth]{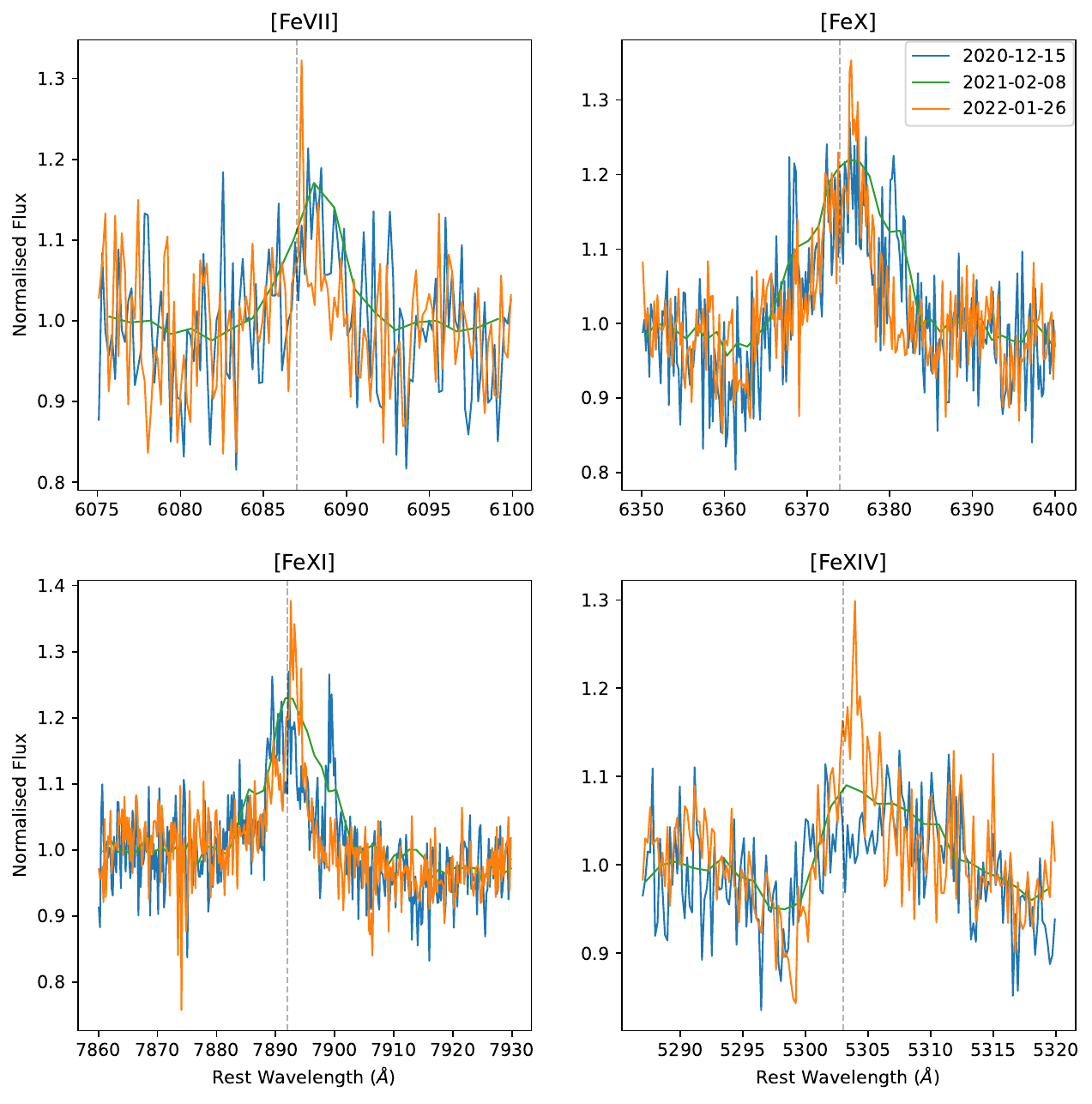}
\caption{Comparison of Fe coronal line regions in all of our late-time spectra. The dotted grey line marks the rest-frame wavelengths of each line. The green spectrum is the MUSE spectrum, the orange and blue spectra are the X-Shooter data.}
\label{fig:fe_comp}
\end{figure*}

In order to determine the source of the narrow line emission we plot our measurements on BPT diagrams \citep{bpt, bpt2, bpt3} in Figure \ref{fig:bpt}. We use the Dec 2020 and Jan 2022 X-Shooter measurements as well as the 2020-02-28 spectrum from \citep{qiz20}. We do not use the Feb 2021 measurement as H$\beta$ is not resolved in the MUSE spectrum. H$\beta$ is difficult to measure in the 2020-02-28 spectrum which gives rise to a large errorbar, however, the [N\,II]/H$\alpha$ and [S\,II]/H$\alpha$ ratios clearly place the galaxy in the AGN region rather than star formation. In the Dec 2020 spectrum the narrow lines look to clearly have an AGN origin, but in the Jan 2022 spectrum the emission looks more like a star forming region. We would expect the ionising continuum for a TDE to be similar to that for an AGN, but any changes in the ionising continuum could cause the line ratios to change. Detailed modelling of TDE emission lines is planned for future work.\\

Using the Jan 2022 spectrum we calculate an He\,II/H$\beta$ ratio of 0.33$\pm$0.05. Assuming both lines are produced via photoionisation, this reflects the relative intensity of the ionising continuum at 912\AA{} and 228\AA{}. If the continuum in this region is a power law then we can calculate the index $\alpha$ using $I_{4686}/I_{4862}\propto(912/228)^{\alpha}$ \citep{penston78}. With our measured ratio we calculate $\alpha=-1.3\pm0.1$. We use this to help construct the SED in Section \ref{sed}.

\subsection{Coronal Lines}
\begin{figure}
\centering
\includegraphics[width=1\linewidth]{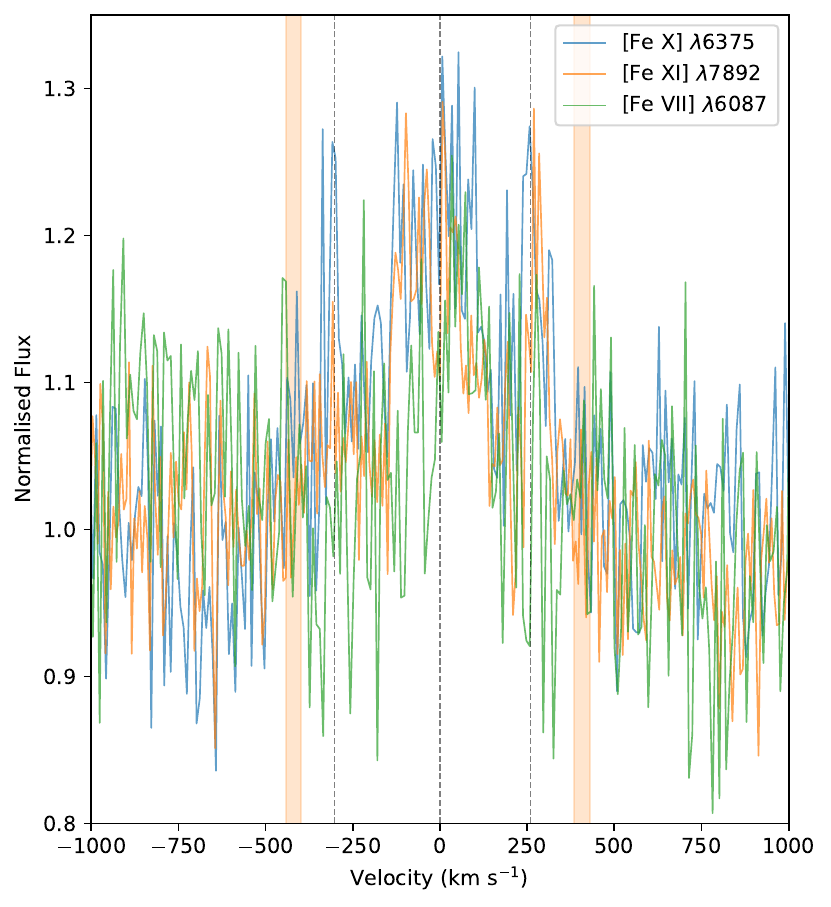}
\caption{Fe lines in the Dec 2020 spectrum overplotted in velocity space. In the [Fe\,X], [Fe\,XI] and maybe [Fe\,VII] regions two features are visible, a broader component centred at 0km/s and narrower line offset by $\sim$260kms$^{-1}$. There is also a feature offset $\sim$300kms$^{-1}$ blueward of [Fe\,X]. The grey dashed lines mark the line centres and the location of the offset components. The orange rectangles mark the locations of broad components fit to H$\alpha$, where the widths cover $\pm1\sigma$. These are offset much more than the narrow Fe features, showing they do not originate from the same region.}
\label{fig:fe_vel}
\end{figure}

The most interesting features in our spectra are the highly ionised Iron lines [Fe\,VII]\,$\lambda$6087, [Fe\,X]\,$\lambda$6375, [Fe\,XI]\,$\lambda$7892 and [Fe\,XIV]\,$\lambda$5304, as well as [Ne\,V]\,$\lambda$3426. In Figure \ref{fig:fe_comp} we display [Fe] line regions from all three of our recent spectra. The most notable change is the appearance of [Fe\,XIV] in the Feb 2021 and Jan 2022 spectra which was not detectable in the Dec 2020 observation. We also note that in the Dec 2020 spectrum narrow features are visible either side of [Fe\,X] and redward of [Fe\,XI]. A feature is also resolved redward of [Fe\,X] in the Feb 2021 spectrum but the feature is not resolved in the [Fe\,XI] region. To determine whether or not these features are related we overplot the Dec 2020 lines in velocity space as shown in Figure \ref{fig:fe_vel}. Here we can see that the red narrow features overlap. The red feature is displaced by $\sim$260kms$^{-1}$ from the line centre, while the blue feature is displaced by $\sim$-300kms$^{-1}$. That the red feature is offset by a similar velocity in both the [Fe X] and [Fe XI] regions suggests that these features are from the same material.
We investigated whether these features lined up with the velocity offsets of the broad Gaussian components fit to the H$\alpha$ blend, but, as shown in Figure \ref{fig:fe_vel}, they clearly originate from a different region.\\

\begin{figure}
\centering
\includegraphics[width=1\linewidth]{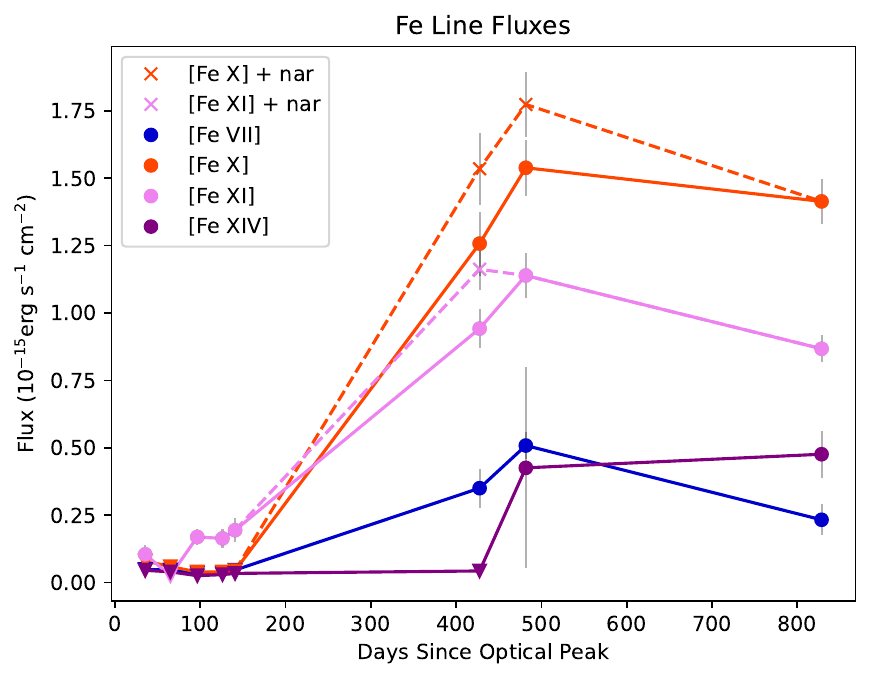}
\caption{Evolution of Fe line fluxes since early outburst spectra. In cases where we don't detect a line we set 1$\sigma$ upper limits denoted by the triangles. We include measurements from the Dec 2020, Feb 2022 and Jan 2021 spectra both with and without the narrow offset features. Measurements with narrow features are marked with crosses and dashed lines.}
\label{fig:fe_fluxes}
\end{figure}

We estimate line fluxes by fitting Gaussian profiles as with the narrow lines above. We fit the offset narrow emission features separately, noting that these are typically unresolved. We also check for the presence of Fe lines in the earlier X-Shooter spectra \citep{qiz20} dating back to $\sim$35 days after lightcurve peak. In most cases the lines are not detected in which case we set an upper flux limit one standard deviation above the continuum flux level. The exception to this is [Fe\,XI] which does seem to appear at earlier times. In Figure \ref{fig:fexi_all} we plot the [Fe\,XI] region for all of the spectra we used. A feature is visible at around the right wavelength in the earlier spectra during outburst, though it is weak, broad and slightly offset so it seems likely this is not real [Fe\,XI] emission. We also plot the rest of the Fe line regions in Appendix \ref{line_plots}. The full list of lines and their flux measurements and line widths is displayed in tables \ref{tab:fe_flx1} and \ref{tab:fe_flx2}. In Figure \ref{fig:fe_fluxes} we plot the evolution of each Fe emission line. In general, the flux of each line increases between the first and second spectra, before decreasing in the most recent spectrum, with the exception of [Fe\,XIV] as it appears for the first time in the Feb 2021 spectrum and increases in the Jan 2022 spectrum. The large errorbar in [Fe\,XIV] flux in the Feb 2021 spectrum comes from the fact that the line is not well resolved by MUSE. However, from Figure \ref{fig:fe_comp} there does appear to be an increase in flux in this region compared to Dec 2020. \cite{wang12} also observe evolution in the Fe line flux of their coronal line emitting objects. They carry out spectroscopic follow-up of three of their targets, taken several years after the first spectrum. They find that the [Fe\,X], [Fe\,XI] and [Fe\,XIV] flux decreases, but that the [Fe\,VII] flux remains constant.\\

The velocity widths of the highly ionised lines range from $\sim$150 to $\sim$300 km s$^{-1}$ , with additional unresolved or barely resolved offset components in [Fe\, X] and [Fe\, XI]. In section \ref{disc} we discuss what this implies for the structure and origin of the illuminated gas in AT2019qiz. Here, as with the low ionisation lines, we examine whether the kinematic information is consistent with the flux evolution. If we once again use a ``virial distance'', we find the coronal line material in the range 50-300 light days. This is qualitatively consistent with the coronal line response peaking well before the low ionisation lines. Quantitatively, it is not quite in simple agreement with observed flux peak at $\sim$500 days, and may suggest somewhat sub-virial velocities, but a proper spatial and kinematic model would be needed before making any stronger statement.

\begin{figure}
\centering
\includegraphics[width=1\linewidth]{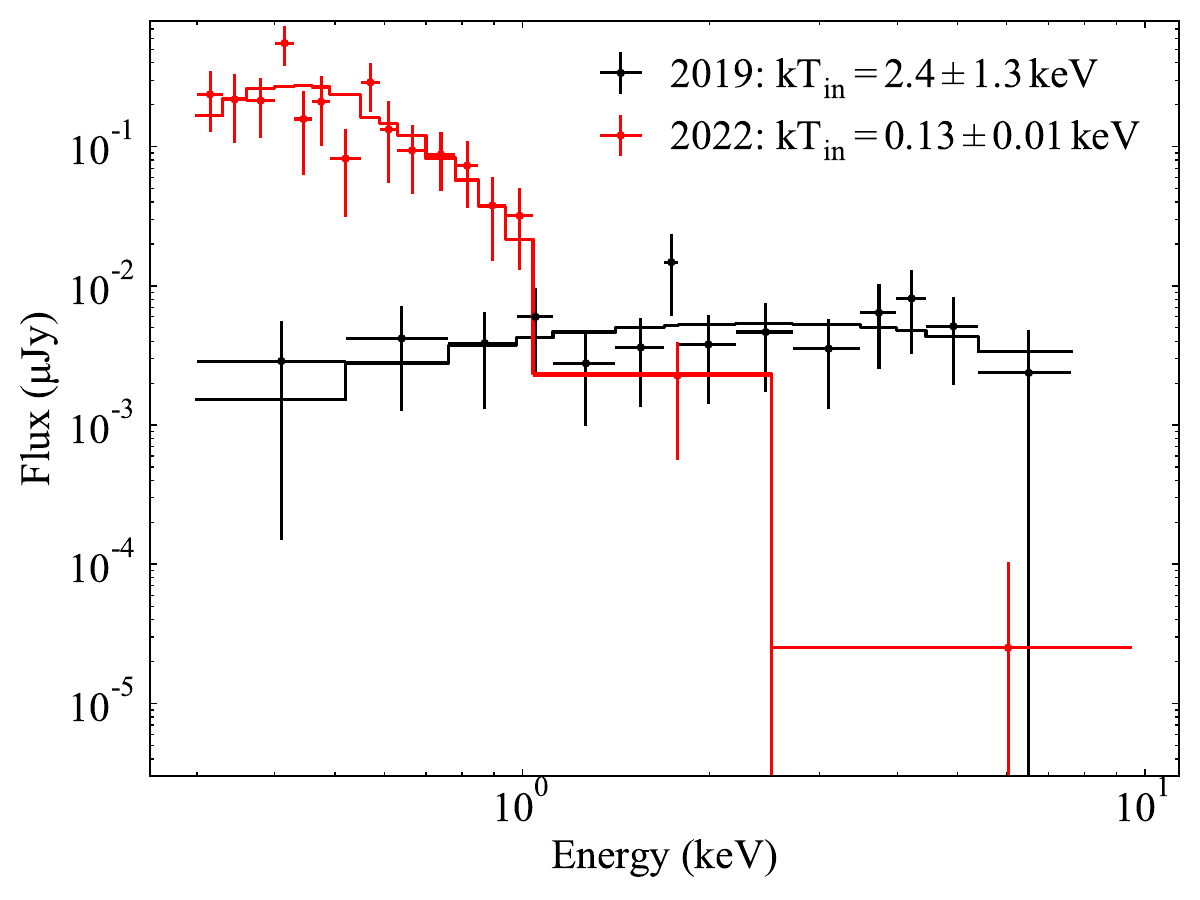}
\caption{Our 2022 XRT spectrum (red) compared to the earlier 2019 observation (black; \citealt{qiz20}). Both spectra can be adequately fit by power-law or thermal disk models (\textsc{diskbb} in \textsc{xspec}). The fit parameters are given in Table \ref{tab:xspec}. The latter models have been used here to convert counts to fluxes, and are shown in the figure labelled by the temperatures at the inner disk edge. The softening of the spectrum is clearly visible.}
\label{fig:xspec}
\end{figure}

\section{X-ray Analysis} \label{swift}

\begin{table*}
    \caption{Parameters and fluxes from X-ray model fits}
    \centering
    \begin{tabular}{cccccc}
         & \multicolumn{2}{c}{Power-law (\textsc{tbabs}$\times$\textsc{zpowerlaw})} & \multicolumn{2}{c}{Thermal disk (\textsc{tbabs}$\times$\textsc{zashift}$\times$\textsc{diskbb})} \\
        Parameter$^a$ & 2019 & 2022  & 2019 & 2022  \\
        \hline
        $\Gamma$ (photon index) & $1.13 \pm 0.26$ & $5.44 \pm 1.12$ \\
        $k_{\rm B}T_{\rm in}$ (keV) & & & $2.54 \pm 1.25$ & $0.13 \pm 0.01$ \\
        Normalization$^b$ & $(8.3\pm2.1)\times10^{-6}$ & $(4.9\pm1.5)\times10^{-5}$ & $(1.2\pm1.9)\times10^{-4}$ & $272\pm151$ \\
        \hline
        \multicolumn{2}{c}{Unabsorbed flux in rest-frame (erg\,cm$^{-2}$\,s$^{-1}$)} \\
        \hline
        $0.3-10$\,keV & $1.1\times10^{-13}$ & $6.9\times10^{-13}$ & $9.4\times10^{-14}$ & $5.2\times10^{-13}$ \\
        $0.3-1$\,keV & $1.0\times10^{-14}$ & $6.6\times10^{-13}$ & $7.4\times10^{-15}$ & $5.1\times10^{-13}$ \\
        $1-6$\,keV & $5.8\times10^{-14}$ & $2.4\times10^{-14}$ & $6.0\times10^{-14}$ & $9.7\times10^{-15}$ \\
    \\
    \end{tabular}
    \\
    $^a$ Redshift fixed at $z=0.0151$ and Galactic column density $n_{\rm H}=6.6\times10^{20}$\,cm$^{-2}$. Any additional intrinsic column density in the host galaxy (modelled using \textsc{ztbabs}) is found to be negligible and unconstrained in both models.\\
    $^b$ Normalization is defined differently for power-law and disk models; see \textsc{xspec} documentation for details.
    \label{tab:xspec}
\end{table*}

\begin{figure*}
\centering
\includegraphics[width=1\linewidth]{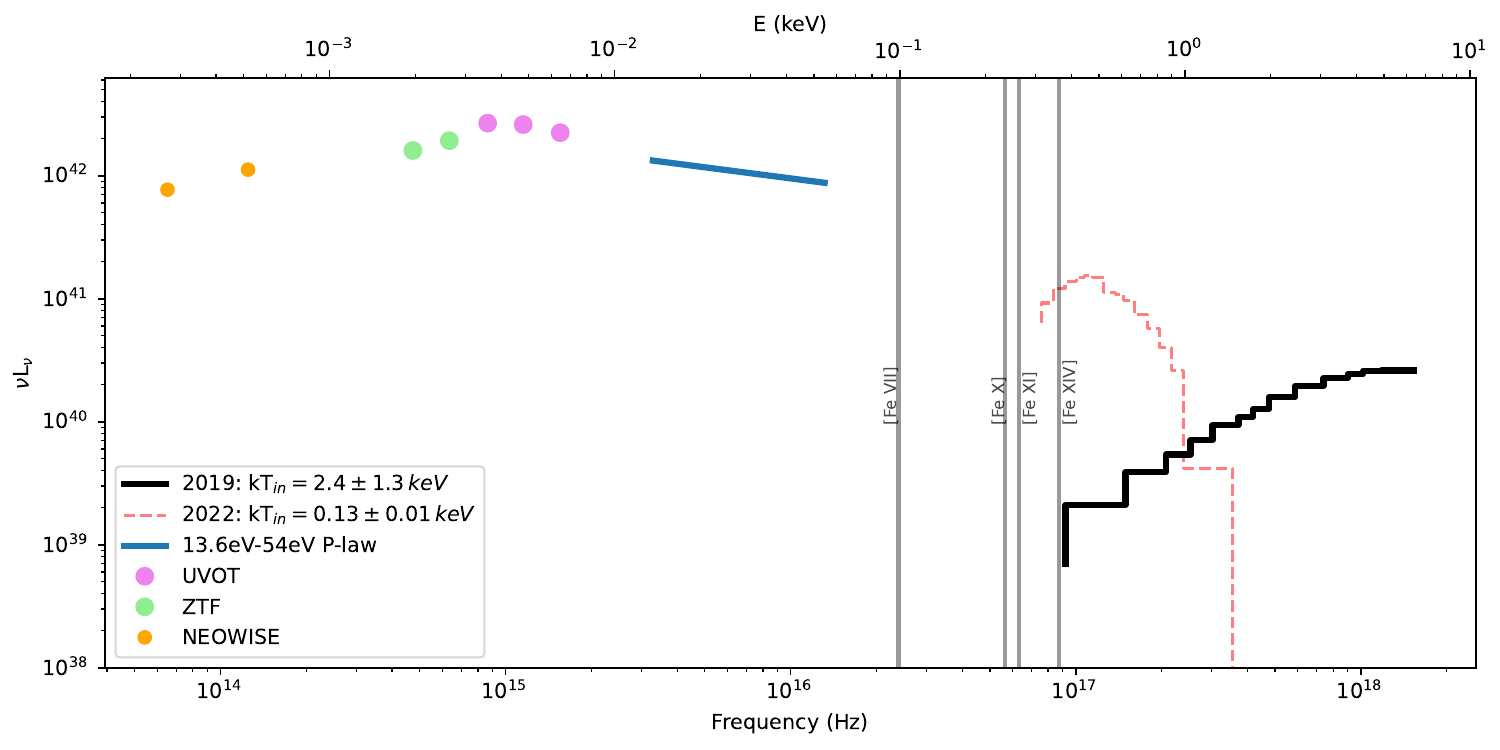}
\caption{The full SED of AT2019qiz in outburst on MJD $\sim$58794. The more recent XRT spectrum is also included to highlight the jump in soft X-ray flux. The IR points are for the first detection on MJD 58886, 92 days after the rest of the SED. Error bars in the photometry points are too small to be seen on this scale.}
\label{fig:SED}
\end{figure*}

AT2019qiz is well detected in our new \textit{Swift} XRT observation. We analyse the X-ray spectrum using \textsc{xspec} \citep{xspec}, and compare it to the stacked XRT spectrum from the time of the optical peak (mean observation date 2019-11-07). We consider two models: a power-law spectrum, and a thermal model consisting of blackbody spectra summed over the annuli of an accretion disk. We initially include both Galactic absorption, determined using the maps from HI4pi \citep{hi4pi} and intrinsic absorption in the host galaxy. However, for all models/epochs, we found an intrinsic column consistent with zero, and therefore exclude this from our final fits for simplicity. In the language of \textsc{xspec}, our models are therefore \textsc{tbabs}$\times$\textsc{zpowerlaw} and \textsc{tbabs}$\times$\textsc{zashift}$\times$\textsc{diskbb}, where \textsc{tbabs} is the  T{\"u}bingen-Boulder ISM absorption model \citep{tubbold}. We fit the data using the Cash statistic \citep{cash79}. The fit parameters and derived fluxes for these models are given in Table \ref{tab:xspec}.\\

Both models provide an adequate fit to the data, with reduced $\chi^2\approx1$. The thermal disk fits are shown compared to the data in Figure \ref{fig:xspec}. Regardless of which model is the more accurate physical description, we find an emphatic change in the spectral slope between the 2019 and 2022 observations. The new data show a much softer spectrum, with an inner disk temperature $\sim 0.1$\,keV (or a photon index $\sim 5$), compared to an almost flat spectrum in 2019. The soft ($0.3-1$\,keV) X-ray flux has increased by more than an order of magnitude, from $\sim 10^{-14}$\,erg\,cm$^{-2}$\,s$^{-1}$ in 2019 to $\gtrsim 5\times 10^{-13}$\,erg\,cm$^{-2}$\,s$^{-1}$ in 2022, while the hard ($1-6$\,keV) X-rays have fallen by a factor of several. We tested whether this could simply occur as a result of a decrease in absorption, by including intrinsic absorption in our models at the redshift of AT2019qiz (\textsc{ztbabs}) and fitting the 2019 data with the temperature or photon index fixed to their best-fit values for the 2022 data. However, this resulted in poor fits, suggesting that the underlying continuum (and not simply the degree of absorption) has changed between 2019 and 2022.

\section{Outburst SED} \label{sed}

\begin{figure*}
\centering
\begin{subfigure}{0.5\textwidth}%
   \includegraphics[width=1\linewidth]{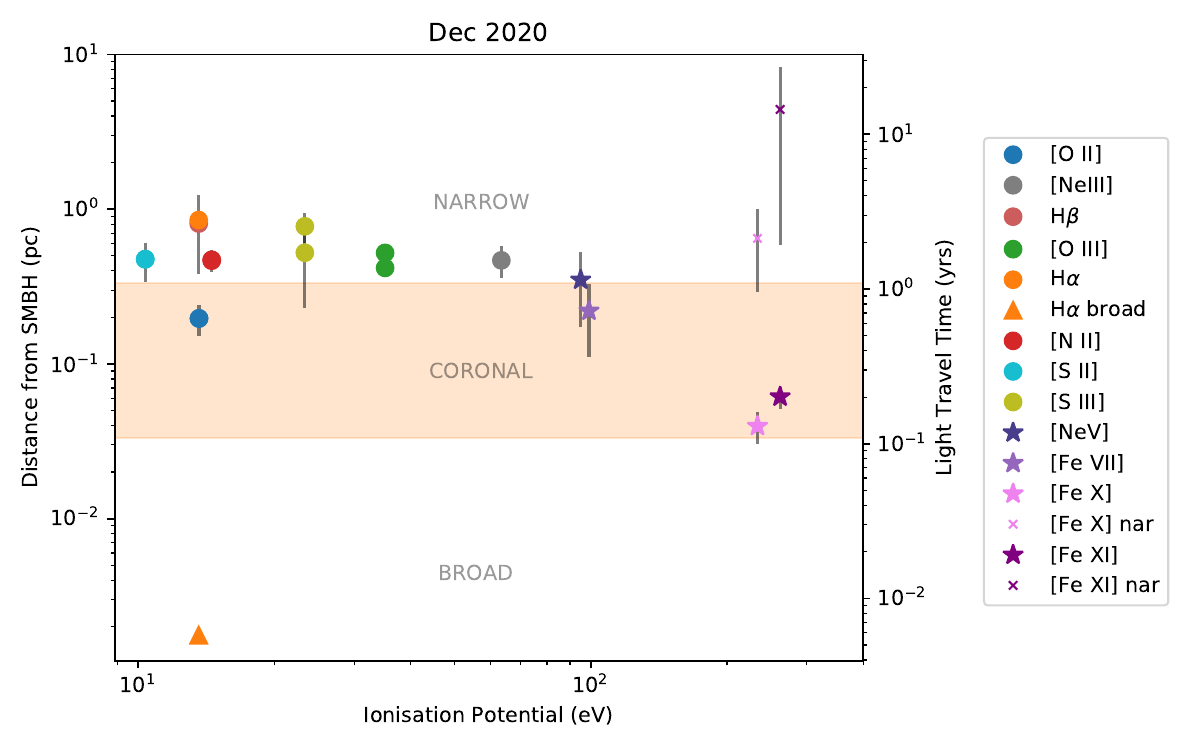}%
   \caption{}%
   \label{fig:ion_1}%
\end{subfigure}%
\begin{subfigure}{0.5\textwidth}
   \includegraphics[width=1\linewidth]{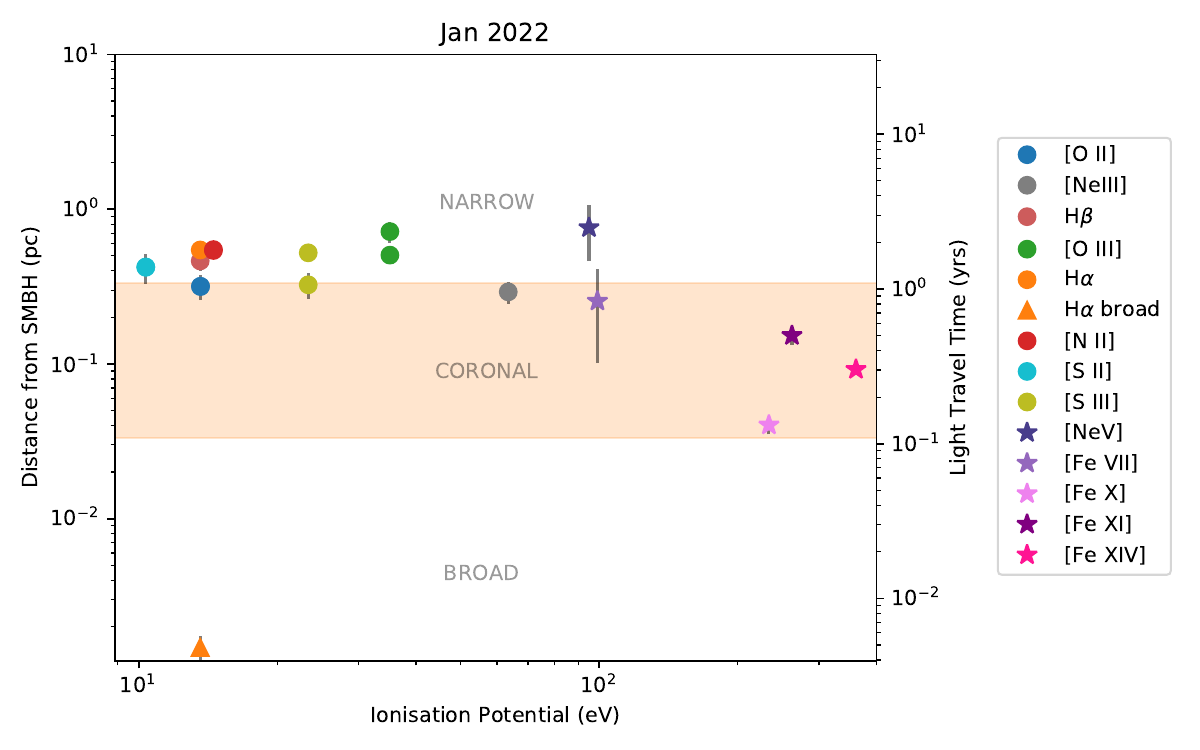}
   \caption{}
   \label{fig:ion_2}
\end{subfigure}
\caption{The estimated distance of the gas emitting various line species, plotted against their ionisation potential. The estimated distance is based on equating the (corrected) FWHM of the line with the expected virial velocity for a  $10^6$M$_{\odot}$ SMBH. Based on these measurements we label narrow, coronal and broad line regions.}
\label{fig:ion}
\end{figure*}

\begin{figure}
\centering
\begin{subfigure}{0.5\textwidth}
   \includegraphics[width=1\linewidth]{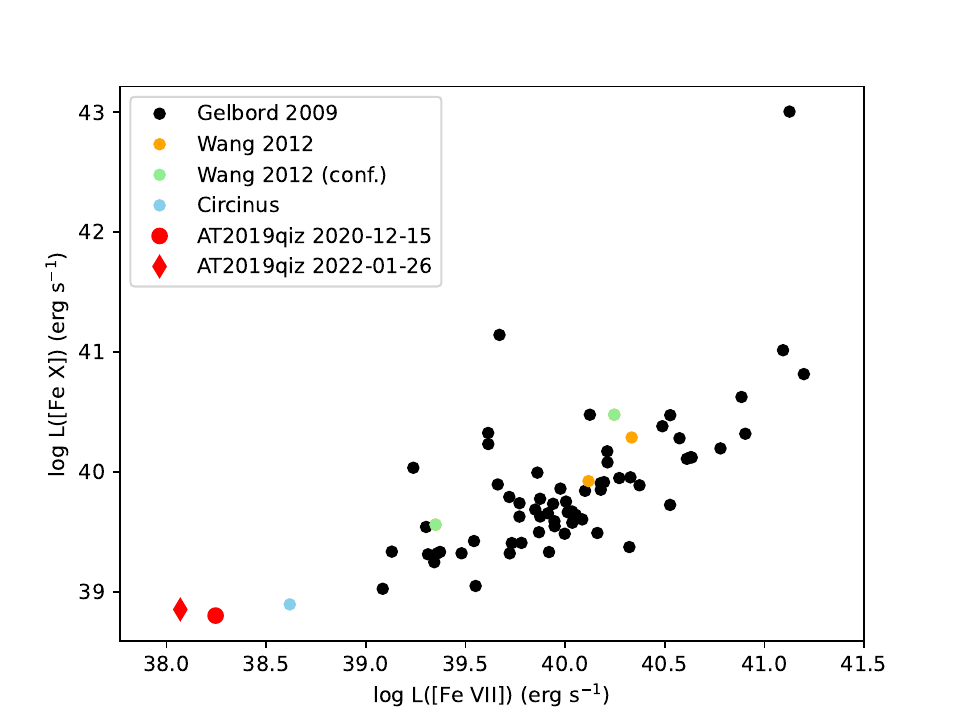}
   \caption{}
   \label{fig:fexfvii} 
\end{subfigure}

\begin{subfigure}{0.5\textwidth}
   \includegraphics[width=1\linewidth]{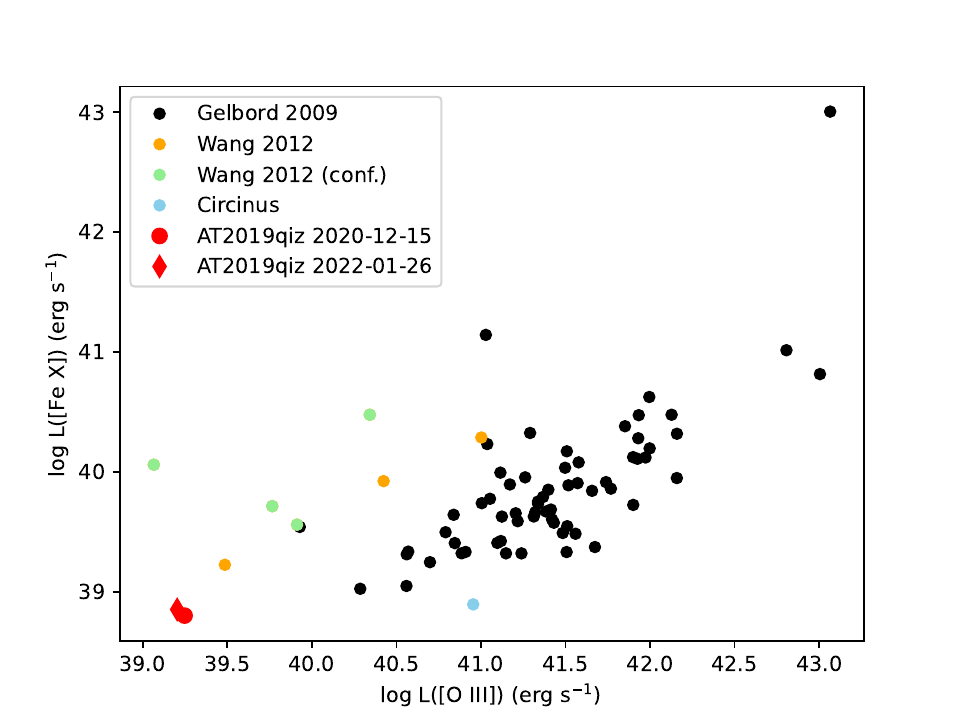}
   \caption{}
   \label{fig:fexoiii}
\end{subfigure}
\caption{A comparison of line ratios with other coronal line emitting objects. The light green points are the four objects in the \citealt{wang12} sample which were confirmed to have variable coronal lines by \citealt{kom08} and \citealt{yang13}. Interestingly, in Figure \ref{fig:fexoiii} AT2019qiz looks more in keeping with suspected TDEs.}
\label{fig:ratios}
\end{figure}

The fluxes of narrow lines at 1-2 light-years from the SMBH reflect the ionising continuum produced 1-2 years earlier, i.e.~the time of the optical peak. This allows us to construct a full SED of the initial flare using constraints from both the early and late-time data, filling in regions that could not be studied in earlier works. In Figure \ref{fig:SED} we attempt to construct the full SED of AT2019qiz during outburst. We use ZTF and UVOT photometry observed around MJD=58794 to coincide with the effective date of the 2019 XRT spectrum. The data, reduced by \cite{qiz20}, were collected from TDE.space\footnote{\href{https://tde.space/}{https://tde.space/}} \citep{guill17}. The infrared (IR) points are NEOWISE data (see Section \ref{IR}) taken 92 days after the rest of the SED. The $13.6-54$eV power law was determined from the He\,II/H$\beta$ ratio as discussed in \S\ref{low-ion} and scaled to be consistent with UVOT photometry. Both the 2019 and 2022 XRT spectra are included in this plot. The 2019 spectrum shows the X-ray flux during outburst and the 2022 spectrum shows how the increase in soft X-ray flux leads to the appearance of the Fe lines, whose ionisation potentials are marked on the plot. Figure \ref{fig:SED} highlights how the photons ionising the coronal lines fill the EUV gap between the He\,II ionising photons and soft X-rays.

\section{Infrared analysis} \label{IR}

\begin{figure}
\centering
\includegraphics[width=1\linewidth]{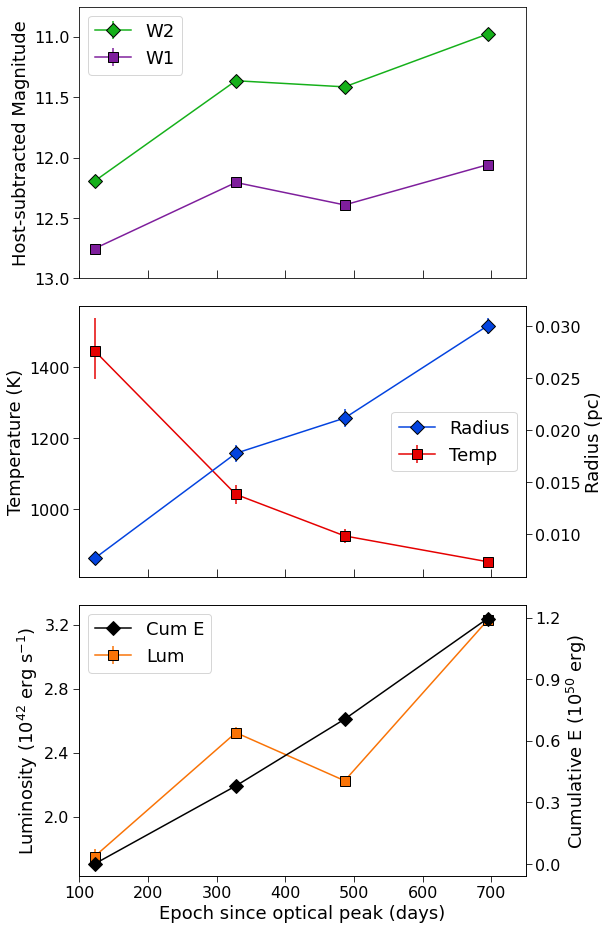}
\caption{Top: Host subtracted magnitudes for NEOWISE observations of AT 2019qiz. Middle: Temperatures and radii inferred from blackbody fitting of the host subtracted fluxes of AT 2019qiz. Bottom: Luminosities implied by the blackbody parameters and the cumulative energy implied by those luminosities.}
\label{fig:WISE}
\end{figure}

AT 2019qiz is well detected in both W1 and W2 bands (3.4$\mu$m and 4.6$\mu$m respectively) for 4 epochs of NEOWISE data. The host is well detected and non-variable in NEOWISE until the first detection of AT 2019qiz at MJD 58886, 122d after the optical peak derived in \citet{qiz20}. The IR fluxes then continue to increase until the most recent observation, 694d after the optical peak at MJD 59458, although there is a slight decline in flux between the 2nd and 3rd observation. We calculate the flux solely due to the transient by subtracting the mean flux value detected from the galaxy before the first detection, and the corresponding magnitudes are shown in the top panel of Fig. \ref{fig:WISE}. The flux at 3.4$\mu$m from the UV/optical blackbody of AT 2019qiz at the epoch of the first IR detection, calculated using the temperature and radius values given in \citet{qiz20}, is $\sim$0.5\% of the observed host-subtracted W1 flux, and so negligible. Therefore, we interpret the IR detections shown here as an IR echo from dust in the vicinity of the SMBH, similarly to those observed in a number of other optical-UV TDEs \citep[see e.g.][]{vanVelzen2016,Jiang2021b}.\\ 

We estimated the BB parameters for the IR echo using the EMCEE python implementation of the Markov Chain Monte Carlo (MCMC) method \citep{ForemanMackey2013} to fit the host-subtracted fluxes and derive uncertainties \citep[as described in][]{Reynolds2022}. The resulting parameters are shown in the middle panel of Fig. \ref{fig:WISE}. The temperature measured in the first observation, T$_{\text{max}}=1450^{+90}_{-80}$~K, is close to, although slightly lower than, the sublimation temperature for astrophysical dust, which typically lies between 1500-2000~K depending on the composition of the dust \citep[see e.g.][]{Waxman2000,vanVelzen2016,2016MNRAS.458..575L}. We note that the inferred radii do not directly correspond to the distance from the SMBH to the dust, and we do not necessarily expect a spherical shell, but do provide a lower limit.\\

We estimate the luminosity of the IR echo at each epoch from the temperature and radius parameters using the Stefan-Boltzmann law, and integrate the luminosity evolution from the first to the last detection to estimate the total radiated energy of the IR echo to date. The luminosity is highest in the most recent observation, where the luminosity is  $3.2\times10^{42}$~erg~s$^{-1}$, and the total radiated energy up until this point is $1.20\times10^{50}$~erg. If we calculate the covering factor by comparing the peak luminosity for the UV/optical blackbody ($3.6\times10^{43}$~erg~s$^{-1}$, \citet{qiz20}) to the peak in the IR, similarly to \citet{Jiang2021b}, we find a lower limit for the covering factor of $f_{c} = L_{\text{dust,peak}} / L_{\text{opt+UV,peak}} = 0.09$, assuming the IR fluxes are still rising. This is considerably higher than the values found for the sample of optical TDEs in \citet{Jiang2021b}, where the covering factor was typically $\sim$0.01. We can additionally consider the definition of the covering factor given in \citet{vanVelzen2021}, $\Omega_{d} = E_{\text{bol}} / E_{\text{dust}}$, where  $E_{\text{bol}}$ and $E_{\text{dust}}$ are the total radiated energy of the TDE (integrated over the wavelength where dust absorption is efficient) and the dust respectively. If we estimate the total radiated energy of the TDE with the value obtained from the UV/optical bolometric LC by \citet{qiz20}, $1.0\times10^{50}$~erg, we find $\Omega_{d} = 0.84$, i.e. an approximate equipartition of the radiated energy between the IR echo and the prompt UV/optical emission. Although the total energy radiated given in \citet{qiz20} only measures until $\sim$150d, if we conservatively assume that the TDE has remained at the same bolometric luminosity from that point until the most recent IR observation, the total radiated energy would only increase by $\sim$20\%, so the value of $\Omega_{d}$ will not be greatly affected. Furthermore, the IR emission is still rising. In summary, the IR echo observed for AT~2019qiz is exceptionally luminous for a TDE discovered in the optical. We discuss this, and the connection to the IR echo observed for AT~2017gge, in Sect. \ref{disc}.

\newpage
\section{Discussion} \label{disc}

We examine the implications of our results for several key issues - the true TDE continuum, the structure and origin of the material surrounding the TDE, and the relation of this object to other coronal line emitters. First however, we discuss whether AT2019qiz was truly a tidal disruption event, as opposed to a supernova, or an AGN outburst.

\subsection{TDE, supernova, or AGN?}

The optical spectrum of the outburst of AT2019qiz fits solidly into the standard class of optical-UV TDEs, but its absolute magnitude at peak, $R_{pk} \sim -19$, was somewhat towards the lower end of the range for such objects \citep{qiz20}, overlapping the range of typical supernovae. At the same time, the low-ionisation line ratios seen in outburst \citep{qiz20} suggest some weak AGN-like activity in the past.

Could AT2019qiz actually be a kind of supernova?  The delayed appearance of coronal lines has been seen in a handful of SNe - for example SN1988z \citep{aretxaga99}, SN2005ip \citep{smith09,smith17}, and SN1987A \citep{groningsson06}; see also discussion in \cite{kom09}.  
Such objects are clearly extremely rare - out of the thousands of supernovae seen so far, only a handful have developed coronal lines. This phenomenon seems to require earlier mass loss to produce significant amounts of circumstellar material; following the outburst, shocks from the blast wave make X-ray emission which then produces the coronal lines. The delay is due to blast wave travel time, not light travel time, and the coronal lines can persist for quite a long time (see references above). 
For the original TDE light echo candidate, J0952+2143, \cite{kom09} were able to argue against a SN origin, because the observed [FeVII] luminosity was  $1.6 \times 10^{40}$ erg s$^{-1}$, two orders of magnitude larger than for these rare extreme SNe.  
For AT2019qiz however, the [FeVII] luminosity is  $3.4 \times 10^{38}$ erg s$^{-1}$, not so far from what was seen in SN2005ip, $1.5 \times 10^{38}$ erg s$^{-1}$ \citep{smith09}. Purely on luminosity grounds, we cannot rule out a supernova origin.  The outburst optical spectrum of AT2019qiz was not quite like any of these extreme SNe - for example, SN2005ip and SN1988z show lines to the red of H$\alpha$ such as the Ca triplet which have never been seen in a TDE. However, supernova spectra are diverse enough that we cannot strictly rule out a supernova origin on this basis either, if we consider AT2019qiz in isolation. However, the similarity of AT2019qiz to other optical-UV TDEs suggests that if we explain AT2019qiz as a supernova, we should consider this possibility for all TDEs, including the more luminous ones. We would also need to explain why these particular characteristics are only found in the centres of galaxies with black holes in the range $10^6 - 10^7 M_\odot$

Could the AT2019qiz outburst be due to some kind of AGN activity, rather than a tidal disruption? In both scenarios we assume the existence of a supermassive black hole, and an enhanced period of accretion, so that the ionising continuum is likely much the same in either case (see section \ref{sec:hard}). The key question  then is the nature and origin of the surrounding material, which we discuss in sections \ref{sec:structure} and \ref{sec:origin}. A lack of pre-existing material would argue for a completely dormant black hole, as opposed to a passive disk around the black hole. However of course, it is quite possible for a stellar disruption to take place in the presence of a passive disk.

\subsection{Existence of a hard continuum}\label{sec:hard}

The coronal lines almost certainly result from photo-ionisation - the very high gas temperatures required mean that collisional ionisation would predict line widths of thousands rather than hundreds of km s$^{-1}$.
The high ionisation coronal lines are then  direct evidence of a hard EUV continuum. While He\,II and Bowen line emission also require an EUV source, the coronal lines have significantly higher ionisation potentials from $\sim$100eV up to nearly 400eV in the case of [Fe\,XIV]. We also find a very soft X-ray spectrum which further hints at the presence of an unobservable continuum. The existence of continuum emission in this energy range is significant as it is predicted by accretion models but is not directly observed. Typical blackbody temperatures observed in TDEs are of order 10$^4$K, which is generally assumed to be reprocessed emission from the true, harder continuum that peaks in the EUV. The detection of coronal lines proves this harder continuum emission exists, and must somehow escape the reprocessing region, perhaps because of either clumpiness, or a non-spherical geometry. In principle, the complete set of emission lines can constrain the shape of the EUV continuum, and the covering factor of illuminated material, but such modelling is beyond the scope of this paper. \\

\subsection{Structure of the surrounding material}\label{sec:structure}

The Fe line light curves in Figure \ref{fig:fe_fluxes} give us the first clues about the geometry of the surrounding structure. The decline of the Fe lightcurve is much slower than that of the continuum, so must be convolved with a broad response function. This can be caused either by  
a spread in light travel times due to a range of radial distances, or due to reflection from material at different angles to our line of sight, or both. Physical models of different reprocessing region geometries produce different response functions (e.g \citealt{perez92}), but unfortunately our sparsely populated light curves make it difficult to compare. The lack of a prompt response to the original outburst tells us there is little material along our line of sight, although another source of uncertainty is whether the coronal lines are responding to the initial outburst or the more recent increase in soft X-ray flux.\\

Inconsistent ionisation parameter changes also show there is a complex structure surrounding the SMBH. [Fe\,XIV] appears in the 2021 and 2022 spectra despite not being observed previously. The other Fe lines decrease in flux after the 2021 spectrum. We also see a significant rise in the flux of narrow lines such as the Balmer lines, [N\,II] and [S\,III]. The changes in the strengths of different emission lines require differing ionisation parameters, which depends on the density of the emitting gas. The later appearance of [Fe\,XIV] could imply low density material further from the SMBH (therefore with a greater light travel time), or alternatively higher density material closer to the SMBH, but behind the nucleus from our point of view.
 Even further from the black hole, there must be material with a low ionisation parameter, thus emitting strongly in H$\alpha$, H$\beta$ and other low ionisation lines. What is not clear is why the narrow Balmer lines get broader in the second spectrum. In addition the move in the BPT diagrams (Figure \ref{fig:bpt}) from obvious AGN emission to something more closely resembling a star forming region is puzzling. Both could be a sign of an increasing contribution to the narrow lines from the TDE emission, causing both a broadening of the narrow lines and altering their ratios by changing the ionising continuum. The jump in broad line flux in the Jan 2021 spectrum could be further evidence for a complex structure. The structure surrounding the black hole may be made up of clumpy material, in which case the obscuration of the central region may fluctuate. If the broad emission line originates from an accretion disk then the increase in flux could be due to there being less material obscuring our line of sight to the disk.\\

The combination of large dust covering factors and coronal Fe lines observed for both AT2019qiz and AT2017gge suggests a connection between the phenomena. Additionally, AT~2019avd also exhibited a very high dust covering factor with f$_{c}\sim$0.5 \citep[][]{Chen2022} along with displaying coronal lines. In particular, the coronal Fe lines could arise from Fe that was locked in dust grains that were destroyed by the UV emission from the TDE. This would be similar to the case of Coronal Line Forest AGNs, where it has been suggested that the coronal lines could be produced in the inner wall of the dusty torus, where the AGN is sublimating the dust \citep{Rose2015}. In the case of AT 2019qiz, both \citet{qiz21} and \citep{qiz20} note that there is likely to be a low luminosity AGN within the host galaxy of AT 2019qiz. The AGN unification model suggests the presence of a dusty obscuring structure, a torus, which here could be responsible for both the IR echo and potentially the coronal lines. The luminosity of the TDE in the UV/optical and lack of evidence for significant extinction implies that the dust producing the IR echo is not substantially in the observer line-of-sight, which could additionally imply an unobscured line of sight to the inner dusty regions that \citet{Rose2015} argue is required to observe the coronal lines. The AGN associated dust structures additionally explain the high covering factor in AT 2019qiz compared to the optical TDE sample, which are almost all in galaxies that do not host an AGN. This is supported by comparison to TDEs and TDE candidates that have been observed in galaxies that host AGN such as Arp 299-B AT1 \citep{Mattila2018,Reynolds2022}, AT 2017gbl \citep[][]{Kool2020} and ASASSN-15lh \citep[][]{Kruhler2018}, which also display very energetic IR echoes, as indeed does the original TDE echo candidate J0952+2143 \citep{kom09}. There is no unambiguous evidence for either AT~2017gge or AT~2019avd hosting an AGN before undergoing their nuclear outbursts, but it also can not be ruled out.\\

Our observations show that in AT2019qiz, the continuum emission must be reprocessed at a range of distances. The Fe line profiles are irregular with disappearing offset components. The observed emission lines also have a range of widths, with the coronal lines generally being broader than the low ionisation narrow lines. We can use our "virial distance" estimates (see \S\ref{low-ion}) to examine the stratification of the material surrounding the outburst event. Figure \ref{fig:ion}  shows estimated distance versus ionisation potential for various observed lines, for each of the two X-Shooter epochs. We can divide material loosely into "broad line", "coronal line", and "narrow line" regions, but even the coronal line region seems to show ionisation stratification. To go beyond this statement would require a full model, beyond the scope of this paper. Furthermore, as discussed in \S\ref{low-ion}, the distance estimates may be an underestimate because the gas motion is responding to the galaxy potential as well as the black hole. The key point however is that the low-ionisation narrow lines are clearly narrower than the coronal lines. \\

\subsection{Origin of the surrounding material}\label{sec:origin}

We have seen that the material causing the IR echo must have a large covering factor and so seems much more likely to be connected with the nuclear activity, rather than being random interstellar material. Assessing the amount of material responsible for the coronal lines is not trivial, as it is rather model independent. Detailed modelling is underway (Yin et al in preparation) but for any reasonable SED,  placing material at the relevant critical density, and at a distance corresponding to the observed delay time, indicates that likewise a significant covering factor is required to reproduce the line flux, of the order 1-10\%. 

Furthermore, the illuminated material is too far from the black hole to have been expelled from the 2019 outburst. We do not have a kinematic model, but the outflow time will likely be of the order of the dynamical timescale $t=\sqrt{R^3/GM}$. If we place the [Fe\, X] material at the light echo distance based on the light curve peak at 500 days, the outflow time is $\sim 4000$ years. The outflow time to the Balmer lines, still rising at 800 days, is at least $\sim 8200$ years. Even at relativistic jet-like outflow velocities, the outflow cannot outpace the illuminating light. The material could have been placed there by previous AGN activity, or the host could have significant amounts of gas along the line of sight. Intriguingly, if we were to speculate that the material is the remnant of a previous TDE, the dynamical timescale would be consistent with the expected average TDE rate of $\sim10^{-4}$yr$^{-1}$galaxy$^{-1}$. Whether or not a past TDE would leave enough material to produce the observed line luminosities, or indeed which of these scenarios is most likely, is left for future work to investigate.\\

\subsection{Comparison to other coronal line emitters}

AT2019qiz and very recently AT2017gge \citep{onori22} are the only spectroscopically confirmed optical-UV TDEs to have developed coronal lines and it is not clear why this is the case. The Dec 2020 spectrum was obtained as part of a study of TDE host galaxies, and was thus one of a sample of other `very late time' TDE spectra. Out of the 11 objects in our sample AT2019qiz was the only one to show signs of Fe lines, despite some of the other host galaxies also hosting AGN. Out of our sample of late time TDE spectra, AT2019qiz was the most recent, and it is also one of the most nearby TDEs discovered to date. It could be that these coronal lines have been produced in other TDEs but simply haven't been detected. However, AT2017gge \citep{onori22} is significantly further away at z=0.0665 and the coronal lines were still detected 1700 days after the initial outburst. It remains unclear what is special about AT2019qiz or AT2017gge. 

Coronal lines are common in AGN \citep{oliva94,gelbord09} 
and of the class of ECLEs, several are strongly suspected  to be light echos from past TDEs \citep{kom08,wang11,wang12}. In Figure \ref{fig:ratios} we compare AT2019qiz with some of these coronal line emitters from the literature on plots of [Fe\,X] vs [Fe\,VII] and [Fe\,X] vs [O\,III]$\lambda$5007. The comparison sample is made up of coronal line AGN from \cite{gelbord09}, a nearby Seyfert 2 (the \textit{Circinus} galaxy) from \cite{oliva94} and the possible TDEs from \cite{wang12} (Note that we have separated the four galaxies (SDSS J0952+2143, SDSS J1241+4426, SDSS J0748+4712 and SDSS J1350+2916) which had confirmed highly variable coronal lines through follow-up spectra). In both plots there is a correlation between the line ratios which is a sign that the lines are photoionised rather than collisionally ionised. In Figure \ref{fig:fexoiii}, the \cite{wang12} objects appear to follow a different correlation to the \cite{gelbord09} AGN. Interestingly, AT2019qiz is more consistent with the \cite{wang12} objects than the AGN, albeit at lower luminosity than all the others. This likely confirms the speculation of \cite{kom08} and \cite{wang12} that the coronal lines in many of these objects are remnants of a TDE outburst. Based on their sample, \cite{wang12} estimate an event rate for coronal line emitters of $\sim10^{-5}$\,galaxy$^{-1}$\,yr$^{-1}$. In our sample of very late time TDE spectra, only $\sim10\%$ of events develop coronal lines. This is consistent with TDE rate estimates \citep{don02, wang04, kesden12}, though we acknowledge the caveats that a) our sample size is small and b) the spectra were obtained at different epochs relative to initial outburst of each event.

\section{Conclusion} \label{conc}
AT2019qiz is the clearest example to date of a spectroscopically confirmed optical-UV TDE developing high ionisation coronal lines in its spectra, appearing $\sim$400 days after the main optical flare. These lines are direct evidence for the presence of a hard EUV continuum. The lines originate from a pre-existing, complex surrounding structure with different ionisation potentials in different regions. Aside from AT2017gge, no other confirmed optical-UV TDEs to date have shown coronal line emission, but the line ratios of AT2019qiz are consistent with coronal line emitters from the literature which are suspected to be TDE remnants. In addition, the estimated rate of coronal line emitters is consistent with the estimated rate of TDEs and the fraction of our late time TDE sample which develop coronal lines. The IR echo observed from AT2019qiz, along with other coronal line emitters, is notably more luminous than typical for optical TDEs, and suggests a connection between these phenomena. This could be explained through the coronal lines arising from Fe liberated from dust grains that were destroyed by the TDE flare. Future studies should systematically search for coronal line emission in late time TDE spectra.

\section*{Acknowledgements}

MN is supported by the European Research Council (ERC) under the European Union’s Horizon 2020 research and innovation programme (grant agreement No.~948381) and by funding from the UK Space Agency. MJW acknowledges support of a Leverhulme Emeritus Fellowship, EM-2021-064, during the preparation of this paper. TMR acknowledges the financial support of the Vilho, Yrj{\"o} and Kalle V{\"a}is{\"a}l{\"a} Foundation of the Finnish academy of Science and Letters. SM acknowledges support from the Academy of Finland project 350458. IA is a CIFAR Azrieli Global Scholar in the Gravity and the Extreme Universe Program and acknowledges support from that program, from the European Research Council (ERC) under the European Union’s Horizon 2020 research and innovation program (grant agreement number 852097), from the Israel Science Foundation (grant number 2752/19), from the United States - Israel Binational Science Foundation (BSF), and from the Israeli Council for Higher Education Alon Fellowship. A. C. Carnall would like to thank the Leverhulme Trust for their support via the Leverhulme Early Career Fellowship scheme. PC and GL are supported by a research grant (19054) from VILLUM FONDEN. MG is supported by the EU Horizon 2020 research and innovation programme under grant agreement No 101004719. FO acknowledges support from MIUR, PRIN 2017 (grant 20179ZF5KS) "The new frontier of the Multi-Messenger Astrophysics: follow-up of electromagnetic transient counterparts of gravitational wave sources" and the support of HORIZON2020: AHEAD2020 grant agreement n.871158. For the purpose of open access, the author has applied a Creative Commons Attribution (CC BY) licence to any Author Accepted Manuscript version arising from this submission.

\section*{Data Availability}

 All spectra will be made available on \href{https://wiserep.weizmann.ac.il/}{WISeREP} upon acceptance of the paper. Swift NEOWISE data are already publicly available.



\bibliographystyle{mnras}
\bibliography{qiz-refs} 


\newpage
\appendix

\onecolumn
\section{Emission Line Fit Results}\label{tables}

\fontsize{8}{10}
\captionsetup{width=17cm}

\begin{center}
\begin{longtable}{ c | c c c }
Line & Centre & Flux & FWHM \\[2pt]
     & \AA{}  & 10$^{-16}$erg\,s$^{-1}$cm$^{-2}$ & km\,$s^{-1}$ \\[2pt]
\hline 
\hline

[Ne\,V]$\lambda$3426 & 3426.27$\pm$0.12 & 5.85$\pm$1.16 & 106$\pm$26\\[2pt]
[O\,II]$\lambda$3727 & 3726.26$\pm$0.15 & 4.46$\pm$0.65 & 144$\pm$16\\[2pt]
[O\,II]$\lambda$3729 & 3729.48$\pm$0.12 & 5.57$\pm$0.68 & 143$\pm$16\\[2pt]
[Ne\,III]$\lambda$3869 & 3869.29$\pm$0.05 & 4.99$\pm$0.44 & 90$\pm$10\\[2pt]
H$\beta$ & 4862.15$\pm$0.09 & 2.44$\pm$0.61 & 65$\pm$16\\[2pt]
[O\,III]$\lambda$4959 & 4959.77$\pm$0.03 & 10.06$\pm$0.41 & 84$\pm$4\\[2pt]
[O\,III]$\lambda$5007 & 5007.59$\pm$0.01 & 35.17$\pm$0.62 & 96$\pm$2\\[2pt]
[Fe\,VII]$\lambda$6087 & 6088.25$\pm$0.28 & 3.5$\pm$0.72 & 143$\pm$33\\[2pt]
[Fe\,X]$\lambda$6375 & 6374.63$\pm$0.27 & 12.57$\pm$1.18 & 330$\pm$38\\[2pt]
[Fe\,X] off & 6380.72$\pm$0.16 & 2.78$\pm$0.63 & 86$\pm$20\\[2pt]
[N\,II]$\lambda$6549 & 6549.74$\pm$0.19 & 2.78$\pm$0.2 & 100$\pm$7\\[2pt]
H$\alpha$ n & 6563.83$\pm$0.03 & 10.69$\pm$0.57 & 77$\pm$4\\[2pt]
H$\alpha$ b & b1 6553.83$\pm$0.46 & 154.43$\pm$7.35 & b1 346$\pm$15\\[2pt]
            & b2 6571.93$\pm$0.49 &                 & b2 313$\pm$20\\[2pt]
[N\,II]$\lambda$6583 & 6584.85$\pm$0.06 & 8.35$\pm$0.61 & 100$\pm$7\\[2pt]
[S\,II]$\lambda$6718 & 6717.66$\pm$0.14 & 2.88$\pm$0.35 & 248$\pm$29\\[2pt]
[S\,II]$\lambda$6732 & 6732.08$\pm$0.18 & 2.23$\pm$0.33 & 246$\pm$0\\[2pt]
[Fe\,XI]$\lambda$7892 & 7891.36$\pm$0.22 & 9.42$\pm$0.73 & 267$\pm$21\\[2pt]
[Fe\,XI] off & 7899.4$\pm$0.08 & 2.2$\pm$0.28 & 43$\pm$8\\[2pt]
[S\,III]$\lambda$9071 & 9070.98$\pm$0.35 & 3.01$\pm$0.69 & 152$\pm$32\\[2pt]
[S\,III]$\lambda$9533 & 9532.54$\pm$0.09 & 5.83$\pm$0.44 & 80$\pm$7\\[2pt]

\caption{Results from Gaussian fits to emission lines in the Dec 2020 spectrum, after correction for instrumental resolution, which is approximately 56 km s$^{-1}$ for blue arm lines, $\lambda <$ 5600 \AA, and  34 km s$^{-1}$ for red arm lines, $\lambda >$ 5600 \AA. As discussed in the text, we estimate that line fluxes have an addition $\sim$ 15\% systematic error.}
\label{tab:fe_flx1}
\end{longtable}
\end{center}


\begin{center}
\begin{longtable}{ c | c c c }
Line & Centre & Flux & FWHM \\[2pt]
     & \AA{}  & 10$^{-16}$erg\,s$^{-1}$cm$^{-2}$ & km\,$s^{-1}$ \\[2pt]
\hline 
\hline

[O\,III]$\lambda$4959 & 4956.88$\pm$0.04 & 16.83$\pm$0.46 & 102$\pm$5\\[2pt]
[O\,III]$\lambda$5007 & 5004.93$\pm$0.02 & 49.64$\pm$0.69 & 109$\pm$3\\[2pt]
[Fe\,XIV]$\lambda$5304 & 5304.02$\pm$1.86 & 4.26$\pm$3.73 & 297$\pm$145\\[2pt]
[Fe\,VII]$\lambda$6087 & 6088.21$\pm$0.16 & 5.08$\pm$0.49 & 125$\pm$19\\[2pt]
[Fe\,X]$\lambda$6375 & 6374.91$\pm$0.19 & 15.38$\pm$1.04 & 300$\pm$27\\[2pt]
[N\,II]$\lambda$6549 & 6550.94$\pm$0.31 & 5.17$\pm$0.47 & 180$\pm$2\\[2pt]
H$\alpha$ n & 6564.36$\pm$0.04 & 16.13$\pm$0.83 & <120$\pm$5\\[2pt]
H$\alpha$ b & b1 6555.83$\pm$0.47 & 228$\pm$10.0 & b1 405$\pm$15\\[2pt]
            & b2 6573.58$\pm$0.33 &                 & b2 255$\pm$18\\[2pt]
[N\,II]$\lambda$6583 & 6585.32$\pm$0.11 & 15.52$\pm$1.4 & 179$\pm$2\\[2pt]
[S\,II]$\lambda$6718 & 6715.1$\pm$0.25 & 3.11$\pm$0.55 & 73$\pm$18\\[2pt]
[S\,II]$\lambda$6732 & 6729.4$\pm$0.2 & 3.81$\pm$0.59 & 73$\pm$18\\[2pt]
[Fe\,XI]$\lambda$7892 & 7892.44$\pm$0.2 & 11.39$\pm$0.84 & 240$\pm$20\\[2pt]

\caption{Results from Gaussian fits to emission lines in the Feb 2021 spectrum,  after correction to instrumental resolution, which varies from 75 to 150 km s$^{-1}$. As discussed in the text, we estimate that line fluxes have an addition $\sim$ 15\% systematic error.}
\label{tab:fe_flx2}
\end{longtable}
\end{center}

\newpage


\begin{center}
\begin{longtable}{ c | c c c }
Line & Centre & Flux & FWHM \\[2pt]
     & \AA{}  & 10$^{-16}$erg\,s$^{-1}$cm$^{-2}$ & km\,$s^{-1}$ \\[2pt]
\hline 
\hline

[Ne\,V]$\lambda$3426 & 3426.26$\pm$0.06 & 4.57$\pm$0.59 & 68$\pm$13\\[2pt]
[O\,II]$\lambda$3727 & 3729.61$\pm$0.08 & 4.46$\pm$0.44 & 112$\pm$10\\[2pt]
[O\,II]$\lambda$3729 & 3726.72$\pm$0.08 & 4.12$\pm$0.43 & 112$\pm$10\\[2pt]
[Ne\,III]$\lambda$3869 & 3869.25$\pm$0.05 & 5.65$\pm$0.38 & 116$\pm$9\\[2pt]
[He\,II]$\lambda$4686 & 4686.62$\pm$0.09 & 2.82$\pm$0.36 & 118$\pm$16\\[2pt]
H$\beta$ & 4862.24$\pm$0.03 & 8.61$\pm$0.86 & 90$\pm$6\\[2pt]
[O\,III]$\lambda$4959 & 4959.41$\pm$0.04 & 8.21$\pm$0.46 & 70$\pm$5\\[2pt]
[O\,III]$\lambda$5007 & 5007.3$\pm$0.02 & 31.8$\pm$0.69 & 86$\pm$2\\[2pt]
[Fe\,XIV]$\lambda$5304 & 5303.71$\pm$0.29 & 4.76$\pm$0.87 & 214$\pm$38\\[2pt]
[Fe\,VII]$\lambda$6087 & 6087.45$\pm$0.31 & 2.33$\pm$0.57 & 107$\pm$37\\[2pt]
[O\,I]$\lambda$6300 & 6301.4$\pm$0.09 & 3.98$\pm$0.5 & 79$\pm$11\\[2pt]
[Fe\,X]$\lambda$6375 & 6374.86$\pm$0.18 & 14.14$\pm$0.85 & 328$\pm$21\\[2pt]
[N\,II]$\lambda$6549 & 6549.67$\pm$0.1 & 4.03$\pm$0.17 & 94$\pm$4\\[2pt]
H$\alpha$ n & 6564.02$\pm$0.01 & 45.37$\pm$0.56 & 94$\pm$1\\[2pt]
H$\alpha$ b & b1 6556.15$\pm$1.45 & 85.44$\pm$15.04 & b1 358$\pm$32\\[2pt]
            & b2 6572.48$\pm$2.38 &                 & b2 363$\pm$60\\[2pt]
[N\,II]$\lambda$6583 & 6584.83$\pm$0.03 & 12.08$\pm$0.52 & 93$\pm$4\\[2pt]
[S\,II]$\lambda$6718 & 6717.84$\pm$0.12 & 3.08$\pm$0.31 & 106$\pm$10\\[2pt]
[S\,II]$\lambda$6732 & 6732.23$\pm$0.14 & 2.72$\pm$0.3 & 105$\pm$10\\[2pt]
[O\,I]$\lambda$7321 & 7321.32$\pm$0.18 & 2.2$\pm$0.37 & 107$\pm$19\\[2pt]
[Fe\,XI]$\lambda$7892 & 7892.85$\pm$0.11 & 8.67$\pm$0.5 & 171$\pm$11\\[2pt]
[S\,III]$\lambda$9071 & 9070.46$\pm$0.12 & 5.58$\pm$0.5 & 119$\pm$10\\[2pt]
[S\,III]$\lambda$9533 & 9532.54$\pm$0.05 & 12.41$\pm$0.5 & 95$\pm$4\\[2pt]

\caption{Results from Gaussian fits to emission lines in the Jan 2022 spectrum, after correction for instrumental resolution, which is approximately 56 km s$^{-1}$ for blue arm lines, $\lambda <$ 5600 \AA, and  34 km s$^{-1}$ for red arm lines, $\lambda >$ 5600 \AA. As discussed in the text, we estimate that line fluxes have an addition $\sim$ 15\% systematic error.}
\label{tab:fe_flx3}
\end{longtable}
\end{center}

\section{Evolution of Iron Coronal lines} \label{line_plots}
In this section we plot coronal line regions from 8 epochs of X-Shooter and MUSE spectra. All plots display a wavelength region ranging from -1500kms$^{-1}$ to 1500kms$^{-1}$ centred on the relevant line. In each case the flux is normalised to the median value of that region.

\begin{figure*}
\centering
\includegraphics[width=0.8\linewidth]{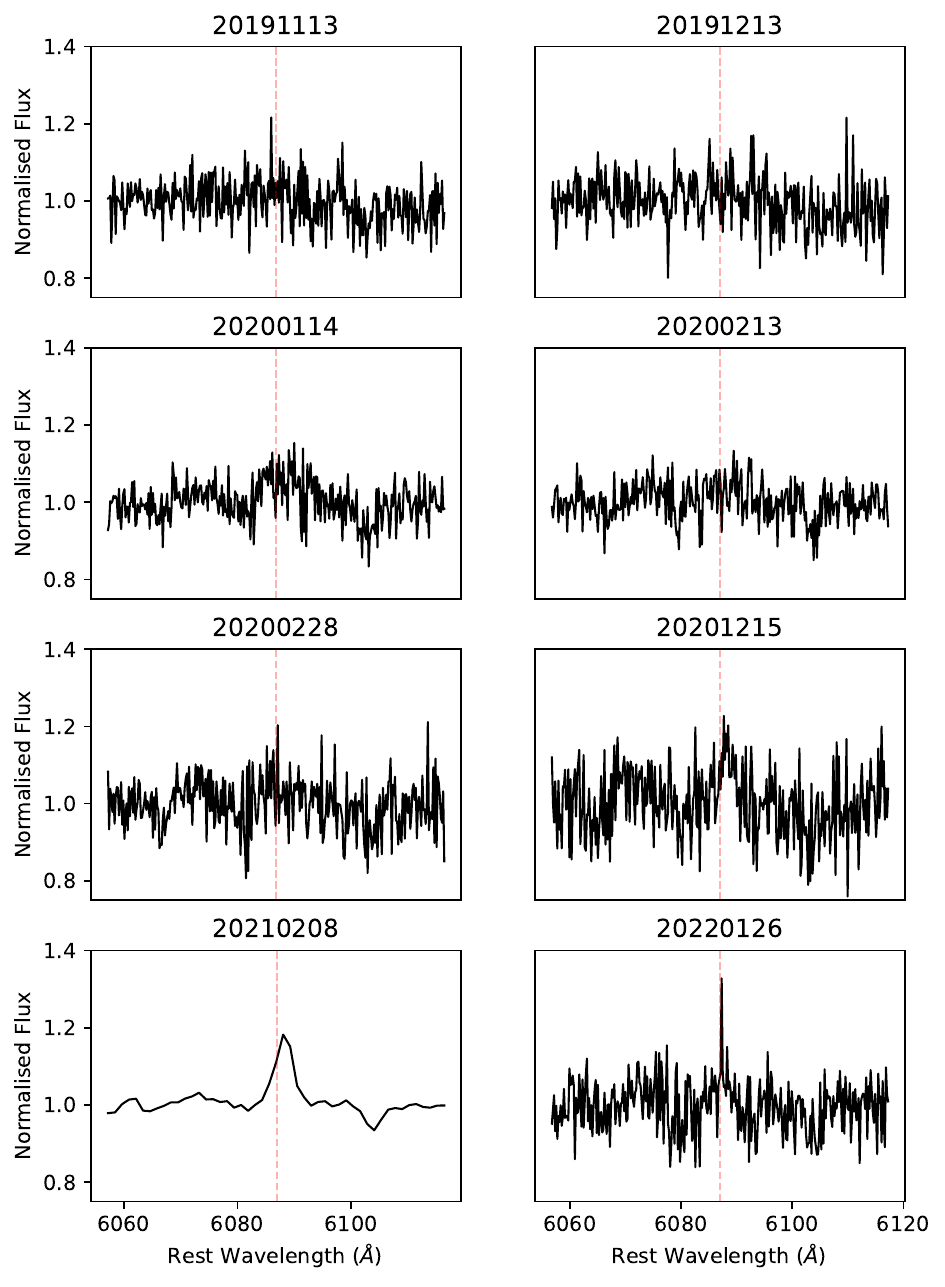}
\caption{Evolution of the [FeVII]\,$\lambda$\,6087 coronal line region.}.
\label{fig:fevii_all}
\end{figure*}

\begin{figure*}
\centering
\includegraphics[width=0.8\linewidth]{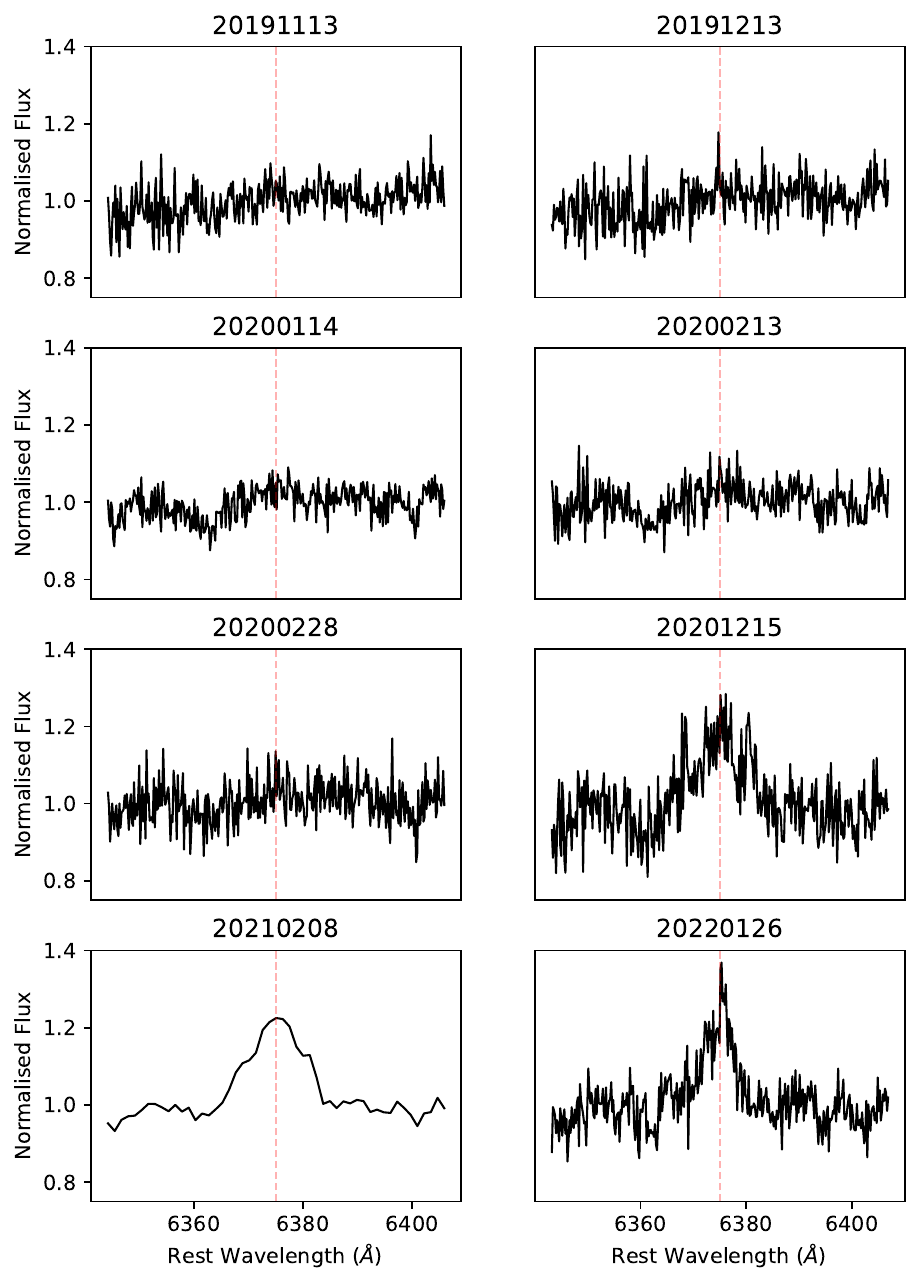}
\caption{Evolution of the [Fe X]\,$\lambda$\,6375 coronal line region.}.
\label{fig:fex_all}
\end{figure*}

\begin{figure*}
\centering
\includegraphics[width=0.8\linewidth]{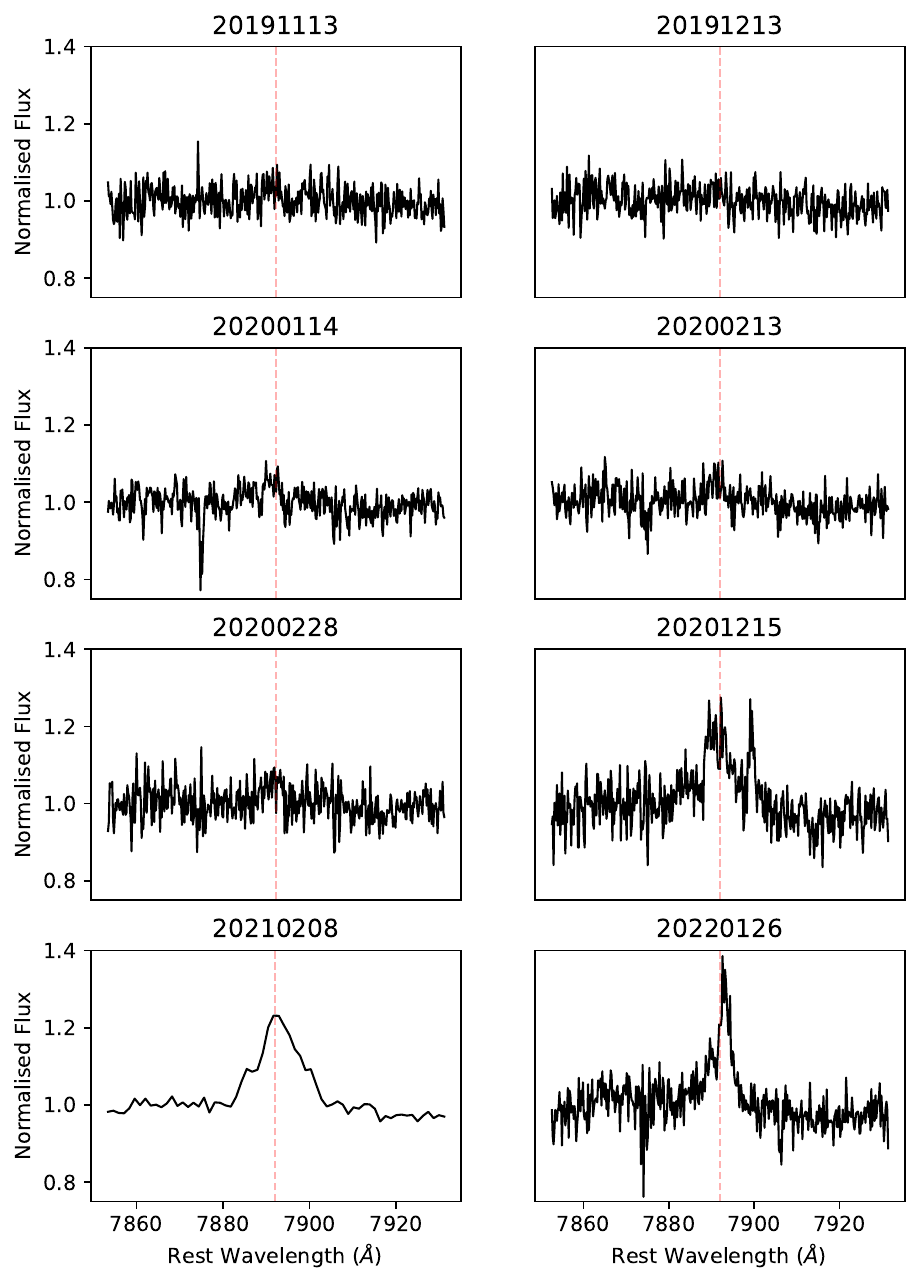}
\caption{Evolution of the [Fe XI]\,$\lambda$\,7892 coronal line region.}.
\label{fig:fexi_all}
\end{figure*}

\begin{figure*}
\centering
\includegraphics[width=0.8\linewidth]{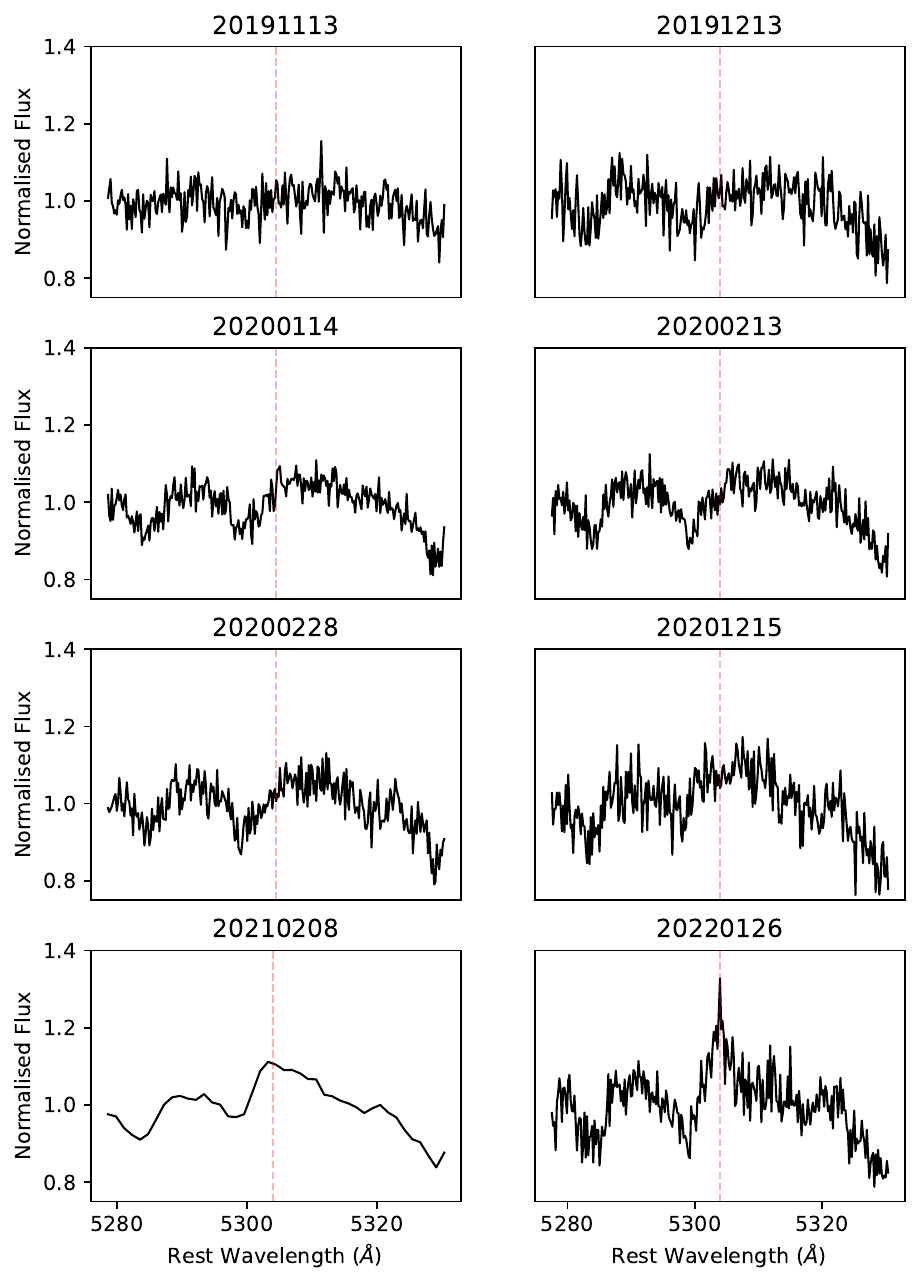}
\caption{Evolution of the [Fe XIV]\,$\lambda$\,5304 coronal line region.}.
\label{fig:fexiv_all}
\end{figure*}


\bsp	
\label{lastpage}
\end{document}